\documentclass[english, eqsecnum, prd, aps, nofootinbib, superscriptaddress, longbibliography, tightenlines,11pt]{revtex4-2}
\usepackage[T1]{fontenc}
\usepackage[utf8]{inputenc}
\usepackage{color}
\usepackage{babel}
\usepackage{mathtools}
\usepackage{amsmath}
\usepackage{amssymb,accents}
\usepackage{mathrsfs}
\usepackage{graphicx}
\usepackage{caption}
\usepackage{subcaption}
\usepackage[unicode=true,pdfusetitle,
 bookmarks=true,bookmarksnumbered=false,bookmarksopen=false,
 breaklinks=true,allcolors=blue,backref=false,colorlinks=true]
 {hyperref}
\usepackage{tikz-cd}

\usepackage{tikz}
\usetikzlibrary{shapes,arrows}

\makeatother
 
\begin{document}

\title{Constraining the quantum gravity polymer scale using LIGO data}

\author{Angel Garcia-Chung}
\email{alechung@tec.mx} 
\affiliation{Tecnol\'ogico de Monterrey, Escuela de Ingenier\'ia y Ciencias, Carr. al Lago de Guadalupe Km. 3.5, Estado de Mexico 52926, Mexico.}
\affiliation{Max Planck Institute for Mathematics in the Sciences Inselstra{\ss}e 22, 04103 Leipzig, Germany}

\author{Matthew F. Carney}
\email{c.matthew@wustl.edu}
\affiliation{Department of Physics and McDonnell Center for the Space Sciences,
Washington University, St. Louis, MO 63130, USA}

\author{James B. Mertens}
\email{jbm120@case.edu}
\affiliation{Case Western Reserve University, Cleveland, OH 44106, USA}
\affiliation{Department of Physics and McDonnell Center for the Space Sciences,
Washington University, St. Louis, MO 63130, USA}

\author{Aliasghar Parvizi}
\email{a.parvizi@ipm.ir}
\affiliation{School of Physics, Institute for Research in Fundamental Sciences (IPM),
P.O. Box 19395-5531, Tehran, Iran}

\author{Saeed Rastgoo}
\email{srastgoo@ualberta.ca}
\affiliation{Department of Physics, University of Alberta, Edmonton, Alberta T6G 2G1, Canada}
\affiliation{Department of Mathematical and Statistical Sciences, University of Alberta, Edmonton, Alberta T6G 2G1, Canada}
\affiliation{Theoretical Physics Institute, University of Alberta, Edmonton, Alberta T6G 2G1, Canada}

\author{Yaser Tavakoli}
\email{yaser.tavakoli@guilan.ac.ir}
\affiliation{Faculty of Physics, University of Warsaw, 
Pasteura 5, 02-093 Warsaw, Poland}
\affiliation{School of Astronomy, Institute for Research in Fundamental Sciences (IPM),	P. O. Box 19395-5531, Tehran, Iran}

\date{\today}
\begin{abstract}
We present the first empirical constraints on the polymer scale describing polymer quantized GWs propagating on a classical background. These constraints are determined from the polymer-induced deviation from the classically predicted propagation speed of GWs. We leverage posterior information on the propagation speed of GWs from two previously reported sources: 1) inter-detector arrival time delays for signals from the LIGO-Virgo Collaboration's first gravitational-wave transient catalog, GWTC1, and 2) from arrival time delays between GW signal GW170817 and its associated gamma-ray burst GRB170817A. For pure-GW constraints, we find relatively uninformative combined constraints of $\nu = 0.96\substack{+0.15 \\ -0.21} \times 10^{-53} \, \rm{kg}^{1/2}$ and $\mu = 0.94\substack{+0.75 \\ -0.20} \times 10^{-48} \, \rm{kg}^{1/2} \cdot s$ at the $90\%$ credible level for the two polymer quantization schemes, where $\nu$ and $\mu$ refer to polymer parameters associated to the polymer quantization schemes of propagating gravitational degrees of freedom. For constraints from GW170817/GRB170817A, we report much more stringent constraints of $\nu_{\mathrm{low}} =2.66\substack{+0.60 \\ -0.10}\times 10^{-56}$, $\nu_{\mathrm{high}} = 2.66\substack{+0.45 \\ -0.10}\times 10^{-56} $ and $\mu_{\mathrm{low}} = 2.84\substack{+0.64 \\ -0.11}\times 10^{-52}$, $\mu_{\mathrm{high}} = 2.76\substack{+0.46 \\ -0.11}\times 10^{-52}$ for both representations of polymer quantization and two choices of spin prior indicated by the subscript. Additionally, we explore the effect of varying the lag between emission of the GW and EM signals in the multimessenger case.
\end{abstract}
\maketitle

\section{Introduction\label{sec:Intro}}

The growing roster of significant gravitational wave (GW) observations continues to provide invaluable insight into the nature of the cosmos \cite{LISA:2022kgy,LISACosmologyWorkingGroup:2022jok,Addazi:2021xuf}. These signals, produced from the collisions of compact objects such as neutron stars and black holes, have profoundly expanded---and continue to expand---the catalog of astrophysical objects in our Universe as well as the properties that describe them. Now the exciting prospect of probing fundamental physics with these signals is upon us, with even more sensitive GW observatories on the horizon \cite{Addazi:2021xuf, LISA:2022kgy, LISACosmologyWorkingGroup:2022jok}. Perhaps one the most enticing prospects of precision GW observations is experimental evidence of the quantum nature of spacetime.

On the other hand, since the formulation of classical general relativity, most of the theory community have a consensus  that gravity, similar to other field of nature, is intrinsically quantum, and thus classical gravity is just a low energy limit of a full theory of quantum gravity. There have been several proposals for such a theory of quantum gravity, none of which are complete as of now. One of the candidates is loop quantum gravity (LQG) \cite{Thiemann:2007pyv, Rovelli:2004tv,Gambini:2011zz} which is a non-perturbative approach proposing quantum states of space are superpositions of gauge-invariant graphs whose edges have labels associated to the gauge group of the theory. The theory is written in terms of a certain connection called the Ashtekar-Barbero connection and the configuration variables are holonomies of this connection over paths in space. 

 The use of holonomies as configuration variables has inspired another model for quantization of both space(time) and matter fields with finite degrees of freedom called the polymer quantization \cite{Ashtekar:2002sn,Tecotl:2015cya,Morales-Tecotl:2016ijb}. This model is closely related to employing Weyl algebra instead of the usual commutation (or Poisson in the classical case) algebra, and hence has at least two representations\footnote{Sometimes also called polarizations.}. Mathematically speaking, this amounts to the use of infinitesimal generators (algebra members) associated to some of the canonical variables, and finite generators (group members) associated to the canonical conjugated variables. Those finite generator mimicking exponentials of infinitesimal generators look very similar to holonomies, and hence this method of quantization looks quite similar to LQG. The dynamics of the polymer approach leads to the quantization/discretization of the canonical variables that are not ``exponentiated'', since their conjugate variables are holonomized or ``exponentiated'' and thus only generate finite transformation of their conjugate non-holonomized counterparts. This method can be applied to both matter, e.g., GRBs \cite{Bonder:2017ckx} and spacetime or its perturbations themselves \cite{Garcia-Chung:2020zyq,Garcia-Chung:2022pdy}.

If gravity or spacetime as a whole is quantized, it implies that its propagating perturbations, i.e., GWs, would also be quantized. When modeling quantum extensions to GWs, there are two approaches: 1) seek out a full quantization of spacetime and derive the quantum-corrected gravitational waveforms as a consequence of the new theory, or 2) adopt a semiclassical approach, separating the GW from its background, and quantize either the background or the GW signal. In this study, we take the latter approach, quantizing the classical transverse-traceless tensor perturbations and assuming that the background spacetime evolves classically. In our previous work \cite{Garcia-Chung:2020zyq}, we derived the equations of motion for polymer-quantized plane wave metric perturbations and provided numerical solutions, as well as approximate analytical solutions, to those equations of motion. Expanding on this work, we have also calculated the alterations to the response of a Michelson-Morley-like GW observatory under the influence of polymer effects \cite{Garcia-Chung:2022pdy, Garcia-Chung:2021doi}. The polymer quantization model has also been applied to the propagation of gamma-ray bursts \cite{Bonder:2017ckx}.

While there have been numerous proposals for tests of the quantum nature of gravity through observations of GWs \cite{Cardoso:2016oxy, Cardoso:2016rao, Abedi:2016hgu, Barcelo:2017lnx, AmelinoCamelia:1997gz}, there has yet to be a test of the polymer quantization as a model for the propagation of GWs. In this work, we provide the first constraints on the polymer scale from GW observations from LIGO's first GW transient catalog (GWTC1) and from the joint detection of the first multimessenger GW event, GW170817 \cite{LIGOScientific:2017vwq}, and its coincident gamma-ray burst signal, GRB170817A \cite{LIGOScientific:2017zic}.

The present paper is structured as follows. In section \ref{theory-review}, we review the procedure for polymer quantizing spacetime metric perturbations using the plane wave approximation, as described in \cite{Garcia-Chung:2020zyq}. We also restate the results for geodesic deviation in orthogonal GW detectors such as LIGO or Virgo. In section \ref{sec:poly-constraints}, we describe how the prediction that polymer effects will cause GWs to propagate slower than classically predicted linearized metric perturbations, can be translated into constraints on the scales introduced during polymer quantization. We outline the process for converting two independent constraints on deviations from classical propagation speed:
\begin{itemize}
 \item[i)] Constraints on $\Delta v_g$  arising from differences in arrival times and distances between detectors.
 \item[ii)] Constraints on $\Delta v_g$  from multimessenger events exhibiting time differences between the GW signal and its associated electromagnetic (EM) counterpart.
\end{itemize}
When applying these methods to events released in the first GW Transient Catalog (GWTC1 \cite{LIGOScientific:2018mvr}), we find that much more informative constraints can be obtained when using multimessenger information on  $\Delta v_g$, subject to the caveat that such constraints are highly sensitive to systematics. Constraints from inter-detector time delays, while much less informative by several orders of magnitude, may be improved with additional GW event data. Finally, in section \ref{sec:conclusion}, we discuss the implications of our results for both approaches and their potential for motivating future studies.

\section{Brief review of the theory\label{theory-review}}
We start from Einstein-Hilbert gravitational action 
\begin{equation}
S_{\rm grav}\ =\ \frac{1}{2\kappa^2}\int d^4x \sqrt{-g}\, \mathcal{R} \, ,
\label{Eq:EH-Action}
\end{equation}
where $\kappa^2\equiv 8\pi G/c^4$, with a general perturbed metric
\begin{equation}
g_{\mu\nu}  =\ \eta_{\mu\nu} +\, h_{\mu\nu}\, ,
\label{Eq:metric-pert}
\end{equation}
in which $\eta_{\mu\nu}$ is the Minkowski metric, 
and $h_{\mu\nu}$ denotes GWs as a small perturbation over $\eta_{\mu\nu}$. We express the perturbation in transverse-traceless gauge as
\begin{equation}
\bar{h}_{\mu\nu}\, :=\, h_{\mu\nu} - \frac{1}{2}\eta_{\mu\nu}h\, ,
\end{equation}
where $h=h^{~\mu}_{\mu}=\eta^{\mu\nu}h_{\mu\nu}$. Given the two polarizations of the GWs and their properties in this gauge, $\bar{h}_{\mu\nu}$
can be expanded explicitly as 
\begin{equation}
\bar{h}_{\mu\nu}=\sum_{\lambda=+,\,\times}\bar{h}_{\lambda}e_{\mu\nu}^{\lambda},
\end{equation}
where $\bar{h}_{+},\,\bar{h}_{\times}$ are the aforementioned two
polarizations of the GW, and 
\begin{align}
e_{\mu\nu}^{+}= & \begin{pmatrix}0 & 0 & 0 & 0\\
0 & 1 & 0 & 0\\
0 & 0 & -1 & 0\\
0 & 0 & 0 & 0
\end{pmatrix}, & e_{\mu\nu}^{\times}= & \begin{pmatrix}0 & 0 & 0 & 0\\
0 & 0 & 1 & 0\\
0 & 1 & 0 & 0\\
0 & 0 & 0 & 0
\end{pmatrix}.
\end{align}
The effective equations of motion  of the independent polarization modes of the waves then reduces to the familiar Klein-Gordon equation,
\begin{equation}
{\Box}\, \check{h}_{\lambda}(x) =0.
\label{Eq:Field}
\end{equation}
where $\check{h}_\lambda(x) \coloneqq \bar{h}_\lambda(x)/ 2\kappa\, $ and $\Box \coloneqq  \eta^{\mu \nu} \partial_\mu \partial_{\nu}$. The conjugate momentum $\check{\pi}_\lambda$ to $\check{h}_\lambda$ is derived as usual using the formula $\check{\pi}_{\lambda}=\frac{\partial{\cal L}_{\check{h}}}{\partial\partial_{t}\check{h}_{\lambda}}$ from the Lagrangian density of the perturbations in terms of $\check{h}_{\lambda}$ given by
\begin{equation}
{\cal L}_{\check{h}}=\frac{1}{2}\sum_{\lambda =+,\times} \check{h}_{\lambda}\mathring{\Box} \check{h}_{\lambda},
\label{eq:Lagrangian-Perturb-new}
\end{equation} 
which is written up to second order in linear perturbations.

In the previous expressions, we have  The classical solutions of the equation (\ref{Eq:Field}) and their conjugates, in Fourier modes are
\begin{subequations}
	\label{eq:H-lambda0}
\begin{align}
\check{h}_{\lambda}(x^0,\mathbf{x})\,  &=\, \frac{1}{\ell^{3/2}}\sum_{\mathbf{k}\in\mathscr{L}}\mathfrak{h}_{\lambda,\mathbf{k}}(x^0)e^{i\mathbf{k}\cdot\mathbf{x}},\label{eq:H-lambda}\\
\check{\pi}_{\lambda}(x^0,\mathbf{x})\, &=\,  \frac{1}{\ell^{3/2}}\sum_{\mathbf{k}\in\mathscr{L}}\Pi_{\lambda,\mathbf{k}}(x^0)e^{i\mathbf{k}\cdot\mathbf{x}},\label{eq: pi-lambda}
\end{align} \label{eq:lambda-tot}
\end{subequations}
where the wave vector $\mathbf{k}=(k_{1},k_{2},k_{3})\in(2\pi\mathbb{Z}/\ell)^{3}$ spans to a three-dimensional lattice $\mathscr{L}$. The canonically conjugate variables $\mathfrak{h}_{\lambda,\mathbf{k}}$ and $\Pi_{\lambda,\mathbf{k}^{\prime}}$ have the Poisson bracket $\{\mathfrak{h}_{\lambda,\mathbf{k}},\Pi_{\lambda,\mathbf{k}^{\prime}}\}=\delta_{\mathbf{k},-\mathbf{k}^{\prime}}$. The reality conditions on the fields indicates that not all the modes are independent. To have an independent expansion for each mode and write the Hamiltonian as a set of decoupled harmonic oscillators, we introduce new new variables ${\cal A}_{\lambda,\mathbf{k}}$
and ${\cal E}_{\mathbf{\lambda,k}}$ and split the lattice $\mathscr{L}$ into positive and negative sectors; for more details see \cite{Garcia-Chung:2022pdy, Garcia-Chung:2020zyq}. In terms of these new variables, the Hamiltonian of the perturbation field reads 
\begin{align}
H\, =\, \frac{1}{2}\sum_{\lambda=+,\times}\sum_{\mathbf{k}\in\mathscr{L}}\left[{\cal E}_{\mathbf{\lambda,k}}^{2}+k^{2}{\cal A}_{\lambda,\mathbf{k}}^{2}\right] \, \eqqcolon\, \sum_{\lambda=+,\times}\sum_{\mathbf{k}\in\mathscr{L}} H_{\lambda, \mathbf{k}},
\label{eq:Hamiltonian-FLRW-2}
\end{align}
where $k=|\mathbf{k}|$.

We now proceed to the polymer quantization of the Hamiltonians in \eqref{eq:Hamiltonian-FLRW-2} in order to extract out the effective terms in the classical limit. To do so, let us first provide the main ideas of the polymer quantization in a self consistent description. To begin with, recall that the fundamental observables required for the Dirac quantization scheme are  ${\cal A}$ and ${\cal E}$ (see \cite{Garcia-Chung:2020zyq} for more details), each for a set $(\lambda, {\bf k})$, and satisfying the Poisson bracket 
\begin{align}
\left\{ {\cal A}_{\lambda, {\bf k}}, {\cal E}_{\lambda', {\bf k}'} \right\} = \delta_{\lambda, \lambda'} \delta_{{\bf k}, {\bf k}},
\end{align}
where $\delta_{\lambda, \lambda'}$ and $ \delta_{{\bf k}, {\bf k}}$ are Kronecker delta functions and the other brackets are null. These Poisson relations are now used to construct the Weyl algebra whose generators, denoted as $W(a,b)$, satisfy the algebra multiplication
\begin{align}
W(a_1,b_1) W(a_2,b_2) = e^{\frac{i}{2} (a_1 b_2 - b_1 a_2)} W(a_1 + a_2,b_1 + b_2).
\end{align}
A better notion of these generators $W(a,b)$ is obtained once we recall that in the standard representation, they can be written as the operator $W(a,b) = e^{i (a {\cal A} + b {\cal E} )}$ and their domain is the entire Hilbert space used in the standard representation. Actually, in the standard representation one can recover the fundamental operators $\widehat{\cal A}$ and $\widehat{\cal E}$ due to the weak continuity condition of the Stone-von Neumann theorem is satisfied. However, polymer quantization violates such a condition and therefore, it is not unitarily equivalent to the standard quantum mechanics \cite{Garcia-Chung:2020zyq,Garcia-Chung:2022pdy}. In this case, the Hilbert space is given by $L^2(\overline{\mathbb{R}},dx_{Bohr})$, where the configuration space, $\overline{\mathbb{R}}$, is the Bohr compactification of the real line and $dx_{Bohr}$ is the Haar measure. Depending on which variable we want to discretize, ${\cal A}$ or ${\cal E}$, the Haar measure will have units of ${\cal E}$ or ${\cal A}$, respectively. We call the first case “polymer ${\cal E}$'' and “polymer ${\cal A}$'' the second case. In the polymer ${\cal E}$ case, the representation is given by
\begin{align}
\widehat{W}(a,b) \Psi({\cal E}) = e^{\frac{i \, a\, b }{2}} \, e^{i b {\cal E}} \, \Psi({\cal E} + a), \qquad \Psi({\cal E}) \in L^2(\overline{\mathbb{R}},d{\cal E}_{Bohr}), \label{polymerE}
\end{align}
and for the polymer ${\cal A}$ the representation takes the form
\begin{align}
\widehat{W}(a,b) \Psi({\cal A}) = e^{- \frac{i \, a\, b }{2}} \, e^{- i a {\cal A}} \, \Psi({\cal A} + b), \qquad \Psi({\cal A}) \in L^2(\overline{\mathbb{R}},d{\cal A}_{Bohr}). \label{polymerA}
\end{align}

The main feature of these representations is of course the discrete spectra of the operators $\widehat{\cal A}$ and $\widehat{\cal E}$ which removes the possibility of having a representation for its canonical partner. That is to say, if $\widehat{\cal A}$ is discrete, then there is no polymer representation for $\widehat{\cal E}$ but for its “exponential'' form $\widehat{W}(0,b)$ and similarly for the case of discrete ${\cal E}$ in which the canonical variable is given in the ``exponential'' form $\widehat{W}(a,0)$.

Depending on the case we are considering, we now impose that every value on the spectra ${\cal A}_j$ (or ${\cal E}_j$) for the discrete operators can be written as
\begin{align}
{\cal A}_j = {\cal A}^{(0)}_j + n \mu, \qquad {\cal A}^{(0)}_j \in [0, \mu) \\
{\cal E}_j = {\cal E}^{(0)}_j + n \nu, \qquad {\cal E}^{(0)}_j \in [0, \nu)
\end{align}
where $\mu$ and $\nu$ are the polymer scales of the system. These polymer scales are considered as the fundamental ``lengths'' for each of the cases. The parameters ${\cal A}^{(0)}_j$ and ${\cal E}^{(0)}_j$ respectively represent the center of the lattice for the polymer states. This can be confirmed by providing an example of polymer state written in the eigenbasis of the discrete operators, also known as almost periodic functions
\begin{align}
\Psi({\cal E}) = \sum_{ \{ {\cal A}_j \}} \Psi_{{\cal A}_j} \, e^{i {\cal A}_j \, {\cal E}} \in L^2(\overline{\mathbb{R}},d{\cal E}_{Bohr}), \\
\Psi({\cal A}) = \sum_{ \{ {\cal E}_j \}} \Psi_{{\cal E}_j} \, e^{i {\cal E}_j \, {\cal A}} \in L^2(\overline{\mathbb{R}},d{\cal A}_{Bohr}).
\end{align}

At this point, we are not interested in the quantum analysis of the Hamiltonians in \eqref{eq:Hamiltonian-FLRW-2} but in their semiclassical versions. To do so, we apply the procedure given in \cite{austrich2017instanton} using path integral analysis. The idea is to obtain the effective action associated with the polymer Hamiltonians using the instanton methods developed for quantum chromodynamic models (for broader polymer examples using group averaging techniques see also \cite{parra2014polymer}). The result yields a modification of the classical Hamiltonians in which the kinetic term is modified in the case of discrete ${\cal A}$ and in the case of discrete ${\cal E}$ the quadratic harmonic potential. We call these modified Hamiltonians effective Hamiltonians\footnote{By effective quantity, e.g., $Q_{\text{eff}}$, we mean an expression of the quantity $Q$ which is derived from the quantum version of that quantity $\hat{Q}$ in a certain way (for example, taking its expectation value of $\hat{Q}$), such that the result is not an operator anymore, but it usually has a modified form compared to the classical version of the quantity. This modification is the result of the process of obtaining the effective  $Q_{\text{eff}}$ (in this example, taking the expectation value of  $\hat{Q}$).}.



 This results in two polymer effective (non-operator) Hamiltonians. For polymer ${\cal E}$ case
Hamiltonian we obtain
\begin{align}
H_{\lambda,{\bf k}}^{({\cal E})}=\frac{2 \hbar^2}{\mu^{2}}\sin^{2}\left(\frac{\mu\,{\cal E}_{\mathbf{\lambda,k}}}{2 \hbar}\right)+\frac{1}{2}{\bf k}^{2}\,{\cal A}_{\mathbf{\lambda,k}}^{2},\label{PolymerE}
\end{align}
and for polymer ${\cal A}$ case the Hamiltonian becomes
\begin{align}
H_{\lambda,{\bf k}}^{({\cal A})}=\frac{1}{2}{\cal E}_{\mathbf{\lambda,k}}^{2}+\frac{2 \hbar^2}{\nu^{2}}\sin^{2}\left(\frac{\nu\,{\cal A}_{\mathbf{\lambda,k}}}{2 \hbar}\right).\label{PolymerA}
\end{align} 
In the plane-wave regime, to the leading order in both  polymer $\cal{A}$ and polymer $\cal{E}$ cases respectively, we obtain the following GW solutions to the equations of motion,
\begin{align}
\bar{h}_{+,\mathbf{k}}^{(\mathcal{E})}(t)\approx & \bar{h}_{I}\left[\left(1-\frac{\bar{h}_{I}^{2}\bar{\mu}^{2}k^{2}}{32\hbar^{2}}\right)\cos\left(kc\sqrt{1-\frac{\bar{h}_{I}^{2}\bar{\mu}^{2}k^{2}}{8\hbar^{2}}}\,t\right)\right.\nonumber \\
 & \quad\quad\quad\quad\left.-\frac{\bar{h}_{I}^{2}\bar{\mu}^{2}k^{2}}{64\hbar^{2}}\cos\left(3kc\sqrt{1-\frac{\bar{h}_{I}^{2}\bar{\mu}^{2}k^{2}}{8\hbar^{2}}}\,t\right)\right],\label{eq:EoM-eff-E-sol-app2}
\end{align} 
and 
\begin{align}
\bar{h}_{+,\mathbf{k}}^{(\mathcal{A})}(t)\, & \approx\,\bar{h}_{I}\left[\left(1-\frac{\bar{h}_{I}^{2}\bar{\nu}^{2}}{96\,\hbar^{2}}\right)\cos\left(kc\sqrt{1-\frac{\bar{h}_{I}^{2}\bar{\nu}^{2}}{8\hbar^{2}}}\,t\right)\right.\nonumber \\
 & \quad\quad\quad\quad\left.-\frac{\bar{h}_{I}^{2}\bar{\nu}^{2}}{192\hbar^{2}}\cos\left(3kc\sqrt{1-\frac{\bar{h}_{I}^{2}\bar{\nu}^{2}}{8\hbar^{2}}}\,t\right)\right].\label{eq:EoM-eff-A-sol-app2}
\end{align}
with the group velocities 
\begin{align}
v^{(\mathcal{A})} \approx c \left(1 - \frac{\bar{h}_I^2 \bar{\nu}^2}{16\hbar^2} \right),  \qquad \qquad v^{(\mathcal{E})} \approx c \left(1 - \frac{3 \bar{h}_I^2 \bar{\mu}^2 }{16\hbar^2}\, k^2 \right). \label{rel:speed}
\end{align}
Here we introduced new polymer parameters 
\begin{align}
\bar{\mu}\coloneqq\mu\ell^{3/2}/2\kappa , \label{eMu} \\
\bar{\nu}\coloneqq\nu\ell^{3/2}/2\kappa, \label{eNu}
\end{align}
\noindent where $\bar{\mu}$ has the dimension of length, and $\bar{\nu}$ is dimensionless in natural units. Depending on the quantization scheme, the velocity is sensitive to either a characteristic polymer length scale, $\bar{\mu}$, or momentum scale, $\bar{\nu}$. The model for the arms of the GW detectors is a system of two free-falling masses. The geodesic separation equation of these masses are sensitive to the metric perturbations $\mathcal{A}_{\lambda, \mathbf{k}}$, i.e., the incident GWs which play the role of a source in this system. The perturbative solutions to the geodesic deviation of the two arms are \cite{Garcia-Chung:2020zyq, Garcia-Chung:2022pdy}
\begin{subequations}
\begin{align}
{\xi}^1_{\bf k}(t) &=  \left[1+ \frac{1}{2} \bar{h}^{(\mathcal{E} / \mathcal{A})}_{+,\mathbf{k}}(t) \right] \xi_0 \cos\theta, \label{rel:xi1Plus1H} \\
{\xi}^2_{\bf k}(t) &=  \left[1 - \frac{1}{2} \bar{h}^{(\mathcal{E} / \mathcal{A})}_{+,\mathbf{k}}(t) \right] \xi_0 \sin\theta ,\label{rel:xi2Plus2H}
\end{align}\label{rel:xiPlus-tot2}
\end{subequations}
\section{Polymer constraints from GW Astronomy\label{sec:poly-constraints}}
In this section, we outline the procedure for leveraging polymer scale-dependent departures from the classical GW propagation speed to place constraints on the polymer scale. The deviation can be inferred from the equations for the group velocity Eq.~\eqref{rel:speed},
\begin{align}
    \label{rel:polyA-dev}
    \Delta v^{\mathcal{A}} &\approx - \frac{c \bar{h}^2_I \bar{\nu}^2}{16 \hbar^2} \\
    \label{rel:polyE-dev}
    \Delta v^{\mathcal{E}} &\approx - \frac{3 c \bar{h}^2_I \bar{\mu}^2}{16 \hbar^2}k^2.
\end{align}
Clearly, Eqs.~\eqref{rel:polyA-dev} and \eqref{rel:polyE-dev} imply any measurements of $\Delta v_g$ can be used to infer constraints on the polymer scales. 

Formally, the extracted probability distribution on the propagation speed, denoted as $p(\Delta v_g)$, is related to the distribution of polymer scales which, we denoted as $p^{\{\mathcal{A},\mathcal{E}\}}(U)$, via a Jacobian transformation. By defining dimensionless parameters $U^{\mathcal{A}} \coloneqq \frac{\bar{h}_I \bar{\nu}}{\hbar}$ and $ U^{\mathcal{E}}(k) \coloneqq \frac{ \bar{h}_I \bar{\mu} }{\hbar}\, k $ in natural units, the relation between the two probabilities distributions, $p^{\{\mathcal{A},\mathcal{E}\}}(U)$ and $p(\Delta v_g)$ becomes 
\begin{equation}\label{jacobi-trans}
    p^{\{\mathcal{A}, \mathcal{E}\}}(U) = \left|\frac{\partial \Delta v_g}{\partial U^{\{\mathcal{A}, \mathcal{E}\}}}\right|p(\Delta v_g).
\end{equation}
 Of course, this relation is valid with the corresponding measures (which we omitted for simplicity) and normalizations of these distributions inside the integrals. 
Invoking Eqs.~\eqref{rel:polyA-dev} and \eqref{rel:polyE-dev} to evaluate Eq.~\eqref{jacobi-trans}, we arrive at a set of simple relations between $p^{\{\mathcal{A}, \mathcal{E}\}}$ and $p(\Delta v_g)$:
\begin{equation}
     p^{\mathcal{A}}(U) = \frac{1}{8} \, U^{\mathcal{A}} \, p(\Delta v_g), \hspace{1 cm} p^{\mathcal{E}}(U(k)) = \frac{ 3}{8} \, U^{\mathcal{E}}(k) \, p(\Delta v_g)
\end{equation}
for the probability distribution for two cases of polymerization.

In the following sections, we use constraints on $\Delta v_g$ from two independent approaches: The first relies on inter-detector arrival time differences for signals detected in multiple GW observatories, while the second compares the arrival time difference between multimessenger GW signals and their electromagntic counterpart. We apply this procedure to event GW170817 and its associated GRB, GRB170817A, as this is so far the only existing confident multimessenger detection.
%
\subsection{Constraints from inter-network arrival time delays}
\subsubsection{Methods}
The following section closely follows section II of \cite{Liu:2020}. In Ref. \cite{Liu:2020}, the propagation speed of the GWs is treated as a free model parameter which deviates from its typical treatment where it is fixed to be the speed of light. The canonical procedure for extracting such parameter information from GWs relies on techniques aimed at sampling the Bayesian posterior probability density.

We use statistical methods to obtain a probability distribution for the inter-detector time delays as a function of $v_g$. This distribution enables us to determine the lower and upper bounds on the speed of gravitational waves. Assuming we have a network of $m$ gravitational wave detectors, each separated by a light travel time $\Delta t^c_{ij}$ (time delay for light between detectors $i$ and $j$), we use the relation$\Delta t ^g_{ij} = c/v_g \Delta t^c_{ij}$ to map light travel time to GW time delay. By considering uniformly distributed sources in the sky and the antenna patterns of the detectors, we can define a distribution of light time delays between every two detectors $p(\Delta t^c_{ij})$. We use this distribution to define the likelihood $p(\Delta t^g_{ij} | v_g)$ \cite{Cornish:2017jml, Liu:2020}. The posterior distribution for $v_g$ for one event and between only two detectors follows from $p(v_g | \Delta t^g_{ij} ) = p(\Delta t^g_{ij} | v_g) p(v_g)/p(\Delta t^g_{ij})$, by the Bayes’ theorem. Here, $p(\Delta t^g_{ij})$ and $p(v_g)$ are the normalization factor and prior knowledge about the distribution in $v_g$, respectively. We assume that $v_g$ follows a uniform distribution.\\
Assuming the data measured at detectors labeled by index $i$ is composed of a signal and noise,
\begin{align}
   d_i(t) = h_i(t) + n_i(t),
\end{align}
where $n_{i}$ is the noise. The probability distribution on this parameter for a single event in a network of detectors can be computed by Bayes theorem:
\begin{align}
    \label{bayes-theorem}
    p(v_g|d_1, d_2, \ldots , d_m) &= \frac{p(v_g)p(d_1, d_2, \ldots , d_m|v_g)}{p(d_1, d_2, \ldots , d_m)}.
\end{align}
The prior distribution $p(v_g)$ encodes any prior knowledge about what values the parameters can take on before a measurement is made. The denominator, $p(d_1,d_2,\ldots, d_m) = \int p(v_g)p(d_1,d_2,\ldots, d_m|v_g)d v_g$, known as the evidence, is a normalization factor useful for model selection that is largely irrelevant for our analysis, and so we do not explicitly compute this. Finally, the likelihood can be written in the frequency domain as
\begin{align}
    p(d_1, d_2,\ldots, d_m|v_g) \propto \prod_i \exp\left[ \int_{-\infty}^{\infty}\frac{|d_i(f)-h_i(f|v_g)|^2}{S_i(f)}df\right],
\end{align}
assuming the noise in each detector, $n_i(t)$, is stationary and Gaussian-distributed, and adopt $d_i(f) = \int_{-\infty}^{\infty}d_i(t)e^{-2\pi i f t}dt$ as our Fourier convention. The remaining components $h_i(f|v_g)$ and $S_i(f)$ are the frequency-domain waveform and power spectral density (PSD) of the noise respectively.  

Lastly, to properly leverage the plethora of GW event data available, Eq.~\eqref{bayes-theorem} can be applied in iteration; for $n$ independent GW events each with data $e_\alpha$, $\alpha = {1, \ldots , n}$, the joint posterior on $v_g$ is proportional to the product of the posteriors from individual events:
\begin{align}
    p(v_g|e_1, e_2, \ldots, e_n) \propto p(v_g|e_1)p(v_g|e_2)\ldots p(v_g|e_n),
\end{align}
assuming a flat prior on $v_g$. Finally, samples drawn from $p(v_g)$ can be trivially converted to samples for the posterior distribution on $\Delta v_g = v_g - 1$ leaving us with $p(\Delta v_g |e_1, e_2 \ldots, e_n)$, which can then be converted to constraints on the polymer scale following the procedure described in section \ref{sec:poly-constraints}.

\subsubsection{Data provenance and modeling}

All posterior samples used in this work were generously provided by the authors of \cite{Liu:2020} in which they used Markov Chain Monte Carlo (MCMC) with Metropolis-Hastings algorithm to sample from multi-dimensional posterior distributions. All binary black hole (BBH) events were modeled using the IMRPhenomPv2 waveform \cite{Khan_2019}, which models the inspiral, merger, and ringdown phases of the GW, and includes the effects of precession on the GW strain. The singular binary neutron star (BNS) event included, GW170817, was modeled using the TaylorF2 waveform, a post-Newtonian inspiral model that includes tidal distortions of the neutron matter.

To apply our model to a binary system we make two simplifying assumptions, which are forced on us from numerical perspectives while on the other hand captures the order of corrections induced by polymerization.  We assume that the source is classical and that it produces GWs with initial amplitude and template for the evolution of the frequency of the inspiral phase and through propagation, the GW waveform receives corrections effectively for each polymer quantization scheme.

Using posterior samples from \cite{Liu:2020} and extracting posterior information on the polymer scale by Jacobian transformation \eqref{jacobi-trans}, we technically assume the true effective waveform (i.e., including polymer corrections) is given by
\begin{align}\label{waveforms-2}
\bar{h}_{+,\mathbf{k}}^{(\mathcal{E})}(t)\approx \bar{h}_{I}\cos\left(kc\sqrt{1-\frac{\bar{h}_{I}^{2}\bar{\mu}^{2}k^{2}}{8\hbar^{2}}}\,t\right)
, \qquad
\bar{h}_{+,\mathbf{k}}^{(\mathcal{A})}(t)\, \approx\,\bar{h}_{I}\cos\left(kc\sqrt{1-\frac{\bar{h}_{I}^{2}\bar{\nu}^{2}}{8\hbar^{2}}}\,t\right),
\end{align}
and that means by analyzing data with classical waveforms such as IMRPhenomPv2 and TaylorF2, we will miss some of the polymer corrections which may induce biases in the recovered posteriors. However, we sidestep these biases since the estimate of $\Delta v_g$ depends solely on the time at which the GW signal amplitude peaks, a model-independent quantity. In a future study we hope to introduce our waveform corrections to LALSimulation and extract the posterior information for polymer parameters directly with polymer-corrected waveforms but this entails a significant project that we leave as a follow-up study. As the final remark, we should note that waveforms \eqref{waveforms-2} are consistent with waveforms employed in templates IMRPhenomPv2 and TaylorF2 while only the speed of GWs is now modified and receives corrections \eqref{rel:polyA-dev} and \eqref{rel:polyE-dev}.
%
\subsubsection{Results}
\begin{figure}
    \centering    \includegraphics[width=\textwidth]{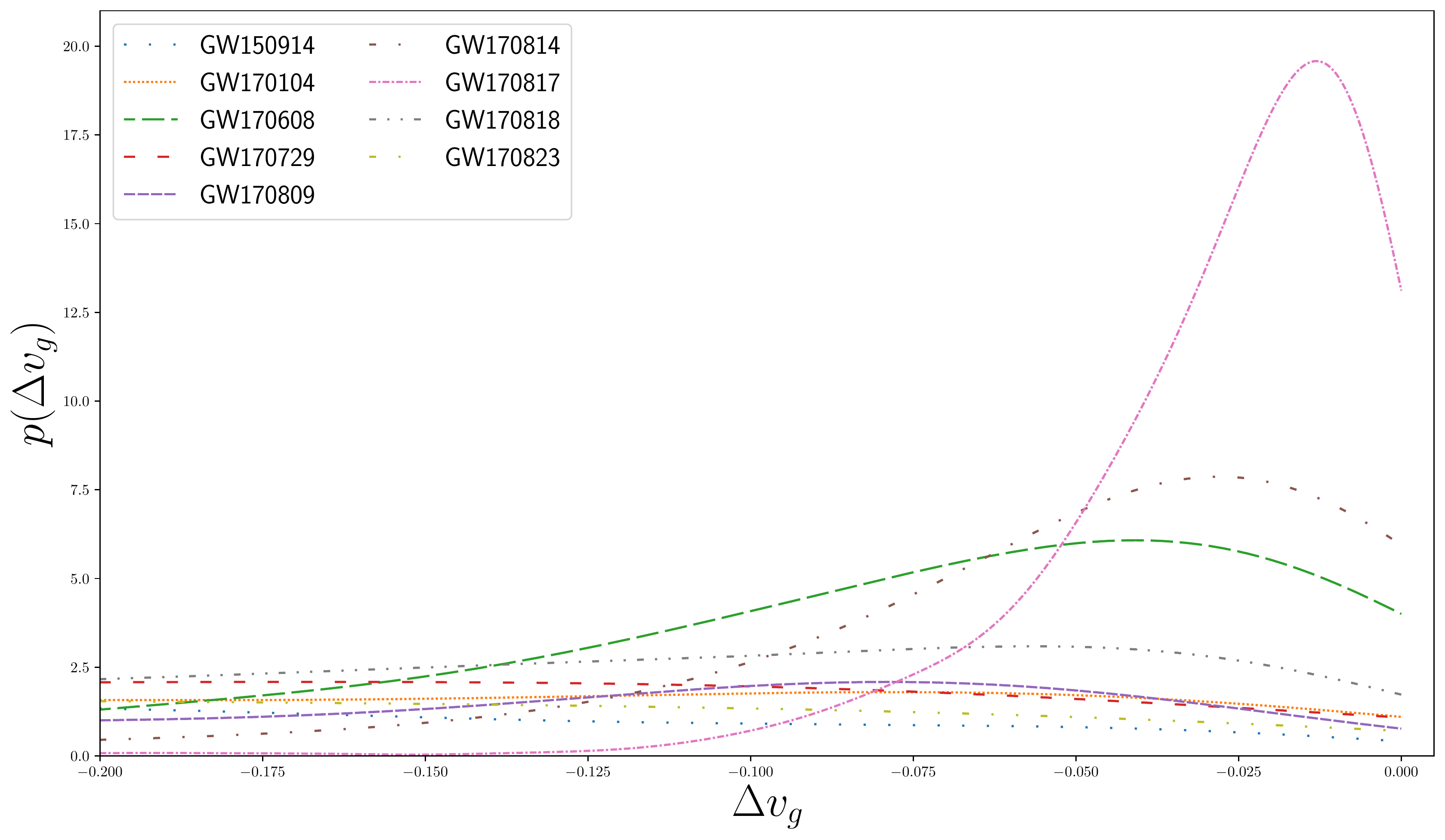}
    \caption{\small Posterior density functions on parameter $v_g$ estimated for events in the first and second observing run of Advanced LIGO (O1 and O2), from which the events with little or no support for negative $\Delta v_{g}$ are removed from the study.}
    \label{fig:O1-O2_prob_vg}
\end{figure}

\begin{figure}
    \centering
\begin{subfigure}[b]{\textwidth}
\includegraphics[width=\textwidth]{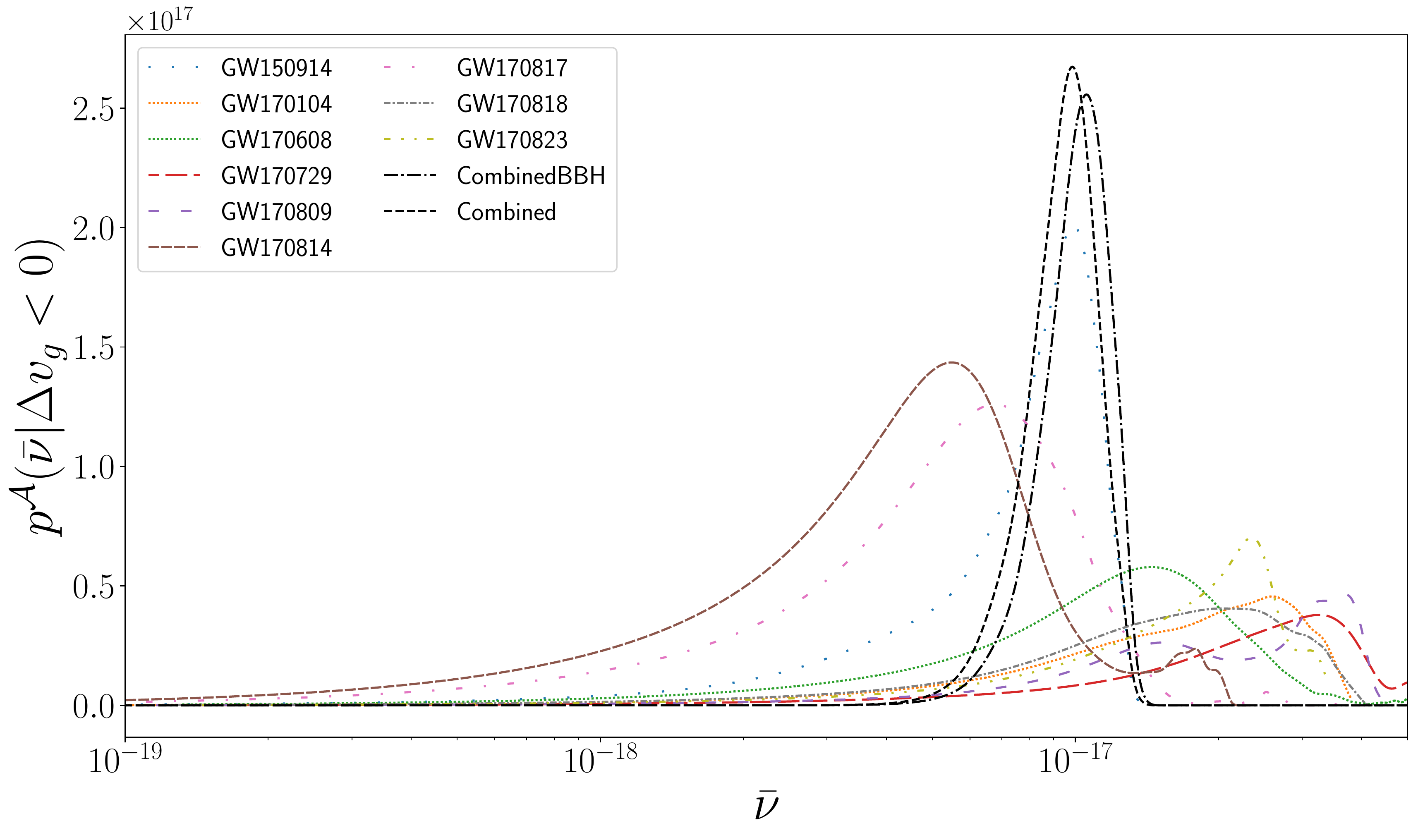}
\caption{Probability density on polymer scale $\bar{\nu}$ given $\Delta v_g < 0$ from gravitational-wave detections from LIGO's first and second observing runs. The combined posterior for $\bar{\nu}$ has a maximum \textit{a posteriori} value of $\bar{\nu}_{\mathrm{MP}} = 0.99\substack{+0.18 \\ -0.22} \times 10^{-17} {\rm kg} \cdot \text{m}^2 \cdot \text{s}^{-1}$. }
\label{fig:polyA_pureGW_constraints}
\end{subfigure}
    
\begin{subfigure}[b]{\textwidth}
\includegraphics[width=\textwidth]{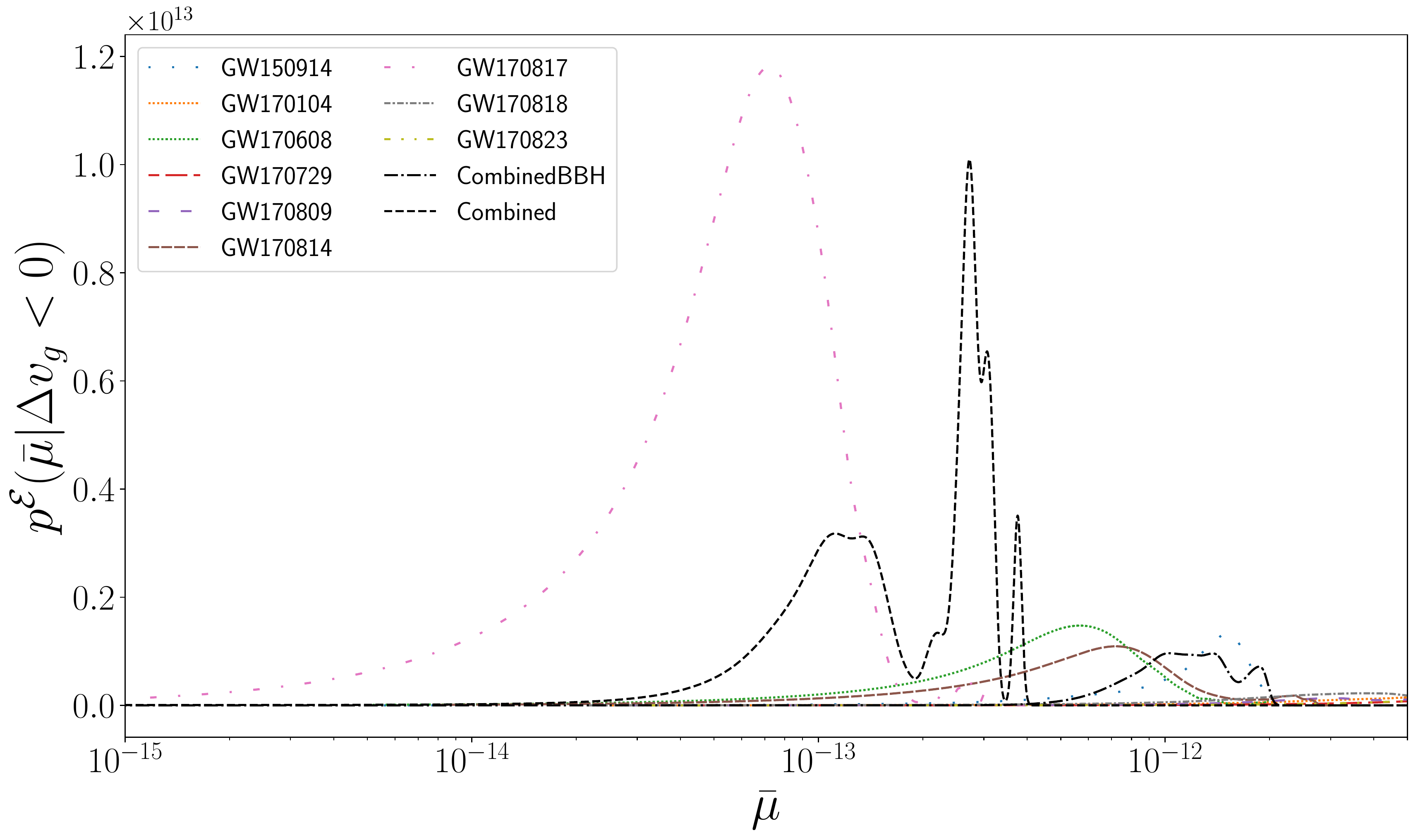}
\caption{Probability density on polymer scale $\bar{\mu}$ given $\Delta v_g < 0$ from gravitational-wave detections from LIGO's first and second observing runs. The combined posterior for $\bar{\mu}$ has a maximum \textit{a posteriori} value of $\bar{\mu}_{\mathrm{MP}} = 0.27\substack{+0.05 \\ -0.17} \times 10^{-12} {\rm kg} \cdot \text{m}^2 $.} \label{fig:polyE_pureGW_constraints}
\end{subfigure}
\caption{\small Posterior probability density functions on polymer parameters extracted from conditional PDF of $\Delta v_g$. All events used for analysis are from the LVC's first gravitational-wave transient catalog paper (GWTC1).}
\end{figure}

Fig.~\ref{fig:O1-O2_prob_vg} displays the probability distribution function (PDF) for $\Delta v_g$ in GWTC1 \cite{LIGOScientific:2018mvr}. The events with little or no support for negative $\Delta v_g$ were excluded from the study. They are composed of 8 BBH events as well as GW170817, the lone BNS event. We combine information constraining $\Delta v_g$  from these 9 events and apply the methods outlined in section \ref{sec:poly-constraints} to resample and interpolate the distribution functions once more. 
Sky localization and high signal-to-noise ratio (SNR) of the events help us better constrain $v_g$, therefore, some events have more sharp peaks. The two highest peaks in Fig.~\ref{fig:O1-O2_prob_vg} correspond to events GW170817 and GW170814, which have high network SNR and their sky location is well constrained. On the other hand, some events like GW170823 and GW170104 have poorly constrained sky localizations and low SNR, thus their posterior plots of $v_g$ seem relatively flat. 
The resulting combined as well as per-event constraints on the polymer scale for each of the two polymer quantization choices are depicted in Figs.~\ref{fig:polyA_pureGW_constraints} and \ref{fig:polyE_pureGW_constraints}. A hard cut prior is placed on all $\Delta v_g$ samples less than 0, as these are excluded by our model. To better present combined results of the events and their probability density distribution on polymer scale $\bar{\mu}$ and $\bar{\nu}$, compute two combined results, in ``CombinedBBH'' only data from BBHs are analyzed, while in ``Combined'' case, data from GW170817, the multi-messenger BNS event is also included in the analysis. Fig.~\ref{fig:polyA_pureGW_constraints} shows that including or removing BNS event data from combined posterior analysis does not affect the results too much, only displacing the maximum of the probability peak by small value, while on the other hand, as presented in \ref{fig:polyE_pureGW_constraints}, BNS data changes the combined analysis considerably. Unsmooth behavior of the combined case for $\bar{\mu}$ is the result of its dependence on the frequency of each signal, we should note that the frequency of BNS event GW170817 is one order larger than the other events. The maximum posterior values of $\Delta v_g$ and polymer parameters $\mu$ and $\nu$ are presented in Table~\ref{tab:peaks}. To find max posterior values for the polymer scales, first, we compute the posterior on $\bar{\nu}$ and $\bar{\mu}$ and then find the maximum of the distribution. In the process, we do not use any of the positive values for $\Delta v_g$ in our analysis, which is why the $y$-axis labels on the polymer scale PDFs now read $p(\bar{\mu} | \Delta v_g < 0)$ as we are effectively computing the conditional PDF. To justify this, we also computed the probability that $\Delta v_g$ is negative for the combined PDFs on $\Delta v_g$, but we have not added them here, which shows more that 50\% of the events. The required length scale $\ell$ for the binary system is set to $10^{10} m$, larger wavelengths are ignored and could be absorbed in the homogeneous background, because we assume our system is localized.

\begin{figure}
    \centering    \includegraphics[scale=0.4]{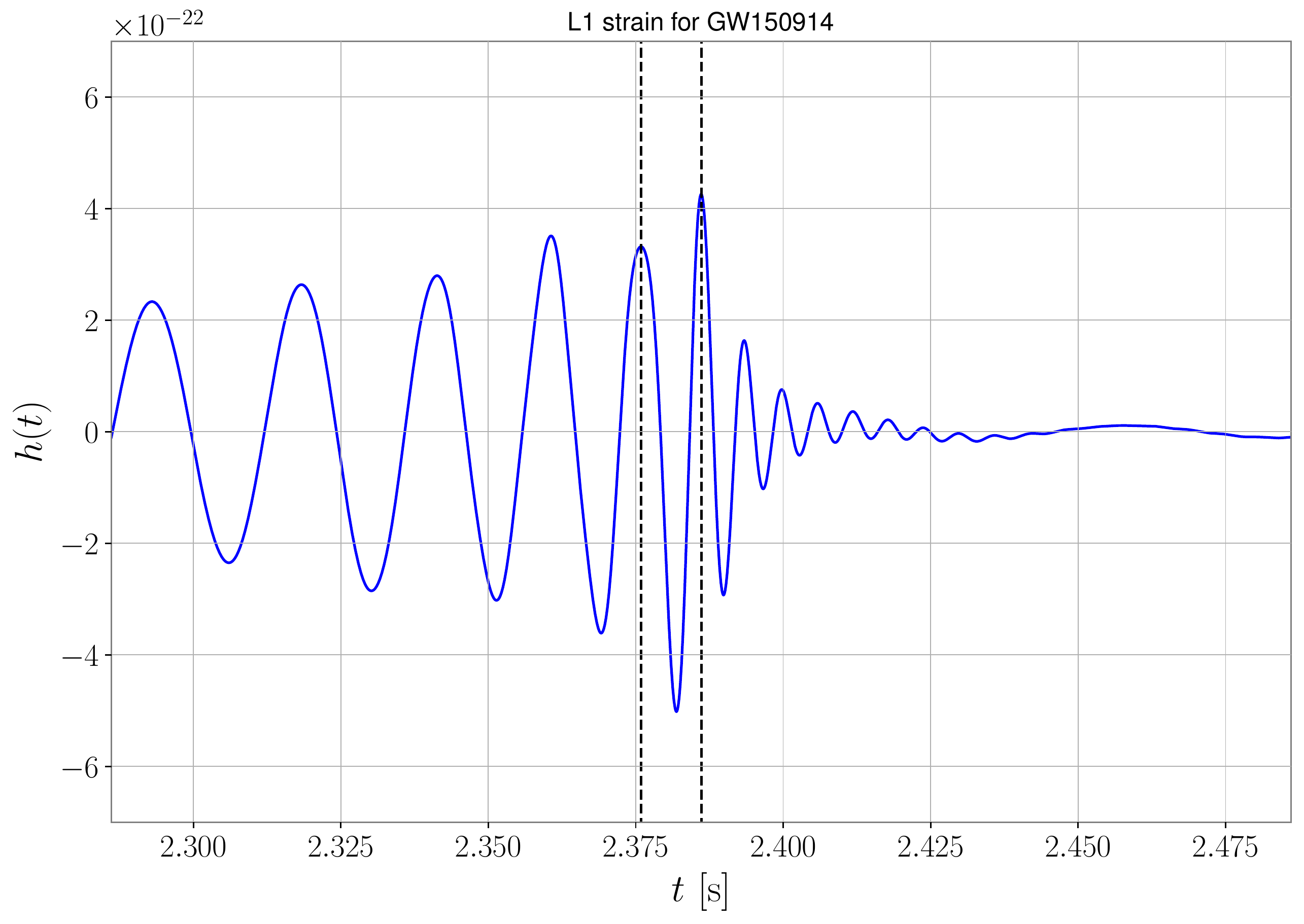}
    \caption{\small Example of a waveform generated to find stain and frequency at the peak. Vertical lines show two consecutive peaks, which are used to extract frequency of GW at the peak.}
    \label{fig:peak}
\end{figure}

We should note here that we did not find any resources that had values for the strain and frequency at the peak for all the BBH events. We instead tried to simulate the time domain signals for each of the events using the maximum posterior values for each of the model parameters. Then, simulated each of the interferometer detector responses which accounts for the antenna function and approximates the noise characteristics using the published power spectral density of the noise for each of the event/detector pairs and found the maximum strain value in each of the detector responses. An example of the generated waveform is plotted in Fig. \ref{fig:peak} for GW150914 event. For the frequency at maximum strain value, we just found the peak just before the merger and did a really rough frequency approximation from the time difference of the two peaks. Our maximum strain value has the same order of magnitude as the few reported values on the available factsheets of the events, but it differs by about $0.8 \times  10^{-22}$. This is within our acceptable range of tolerance, because by assuming waveforms \eqref{waveforms-2}, we have already accounted for this level of uncertainty. After all, in the combined cases and corresponding values for polymer scales, the effect of these small tolerances will get even smaller.  
%
%
\begin{table}
\scriptsize
  \centering
\begin{tabular}{p{.1\textwidth}p{.15\textwidth}p{.1\textwidth}p{.09\textwidth}p{0.09\textwidth}p{0.09\textwidth}p{0.09\textwidth}p{0.09\textwidth}p{0.05\textwidth}}
\hline \hline
Events &  $\max(\Delta v_g)$ & $\bar{\nu}(/10^{-17})$ $\text{kg}\cdot \text{m}^2/\text{s}$ & $\nu (/10^{-53})$ $\text{kg}^{1/2}$ & $\bar{\mu}(/10^{-12})$ $\text{kg} \cdot \text{m}^2$ & $\mu(/10^{-48})$ $\text{kg}^{1/2} \cdot \text{s}$ & GW strain at Peak & frequency at peak Hz & SNR \\[1ex]
\hline
GW150914 & $-0.39\substack{+0.48 \\ -0.16}$ & $0.99\substack{+0.16 \\ -0.45}$ & $0.9\substack{+0.15 \\ -0.41}$ & $1.50\substack{+0.25 \\ -0.68}$ & $1.37\substack{+0.23 \\ -0.62}$ & $1.6\times 10^{-21}$ & 181 & 24.4 \\[1ex]
GW170104 & $0.23\substack{+1.90 \\ -0.54}$ & $2.60\substack{+0.64 \\ -1.56}$ & $2.37\substack{+0.58 \\ -1.42}$ & $6.63\substack{+1.63 \\ -0.40}$ & $6.05\substack{+1.49 \\ -3.63}$ & $5.9\times 10^{-22}$ & 108 & 13.0 \\[1ex]
GW170608 & $0.88\substack{+2.40 \\ -1.50}\times 10^{-1}$ & $1.45\substack{+1.19 \\ -0.70}$ & $1.32\substack{+1.09 \\ -0.63}$ & $0.57\substack{+0.47 \\ -0.27}$ & $0.52\substack{+0.43 \\ -0.25}$ & $4.4\times 10^{-22}$ & 702 & 14.9 \\[1ex]
GW170729 & $3.13\substack{+1.35 \\ -2.00}$ & $3.26\substack{+1.05 \\ -1.73}$ & $2.97\substack{+0.96 \\ -1.58}$ & $8.95\substack{+2.89 \\ -4.76}$ & $8.15\substack{+2.64 \\ -4.34}$ & $4.1\times 10^{-22}$ & 100 & 10.8 \\[1ex]
GW170809 & $-0.79\substack{+0.45 \\ -5.61}\times 10^{-1}$ & $3.72\substack{+0.17 \\ -2.47}$ & $3.39\substack{+0.16 \\ -2.24}$ & $7.28\substack{+0.34 \\ -4.85}$ & $6.64\substack{+0.31 \\ -4.42}$ & $5.0\times 10^{-22}$ & 141 & 12.4 \\[1ex]
GW170814 & $0.17\substack{+0.51 \\ -1.25}\times 10^{-1}$ & $0.55\substack{+0.99 \\ -0.26}$ & $0.50\substack{+0.90 \\ -0.24}$ & $0.72\substack{+1.30 \\ -0.35}$ & $0.66\substack{+1.17 \\ -0.32}$ & $9.6\times 10^{-22}$ & 210 & 15.9 \\[1ex]
GW170817 & $1.88\substack{+2.68 \\ -5.53}\times 10^{-2}$ & $0.68\substack{+0.0.53 \\ -0.33}$ & $0.62\substack{+0.48 \\ -0.30}$ & $0.07\substack{+0.06 \\ -0.03}$ & $0.07\substack{+0.05 \\ -0.03}$ & $5.3\times 10^{-22}$ & 2582 & 33.0 \\[1ex]
GW170818 & $0.51\substack{+0.89 \\ -4.20}\times 10^{-1}$ & $2.04\substack{+1.23 \\ -1.06}$ & $1.86\substack{+1.12 \\ -0.96}$ & $3.69\substack{+2.22 \\ -1.91}$ & $3.36\substack{+2.03 \\ -1.74}$ & $4.9\times 10^{-22}$ & 152 & 11.3 \\[1ex]
GW170823 & $1.96\substack{+8.73 \\ -1.41}$ & $2.35\substack{+0.54 \\ -1.24}$ & $2.14\substack{+0.50 \\ -1.13}$ & $8.76\substack{+2.02 \\ -4.64}$ & $7.99\substack{+1.84\\ -4.23}$ & $6.6\times 10^{-22}$ & 74 & 11.5 \\[1ex]
\hline \hline
Combined\newline(BBH) & $0.09\substack{+5.73 \\ -5.45}\times 10^{-2}$ & $1.06\substack{+0.17 \\ -0.23}$ & $0.96\substack{+0.15 \\ -0.21}$ & $1.03\substack{+0.82 \\ -0.21}$ & $0.94\substack{+0.75 \\ -0.20}$ &  &  & \\[1ex]
Combined & $1.30\substack{+2.76 \\ -3.63}\times 10^{-2}$ & $0.99\substack{+0.18 \\ -0.22}$ & $0.90\substack{+0.16 \\ -0.20}$ & $0.27\substack{+0.05 \\ -0.17}$ & $0.25\substack{+0.05 \\ -0.15}$ &  &  & \\[1ex]
\hline
\end{tabular}
\caption{\small Locations of maximum a posteriori values of $\Delta v_g$, $\bar{\nu}$ and $\bar{\mu}$ for all the events, and the corresponding calculated polymer parameters in their reduced form $\nu$ and $\mu$. ``Combined'' and ``Combined (BBH)'' refer to combined events data with/out the BNS event GW170817. Uncertainties listed are calculated to the 90\% credible level. To have a better upper bound estimates for the polymer parameters, we use the frequency and strain of the peak of inspiral phase with tolerance about $0.8 \times  10^{-22}$, where we assumed $\ell  = 10^{10}\,\text{m}$ for the length scale of the system.}
\label{tab:peaks}
\end{table}
%
%
\subsection{Multi-messenger constraints}
Multi-messenger astronomy has developed rapidly over the past years. The channel type of astronomical messengers now includes electromagnetic radiation, gravitational waves, neutrinos and cosmic rays. One of the main multi-messenger sources are binary pairs (BHs and NSs) \cite{Sana:2012px}, since their first detections in 2015 by LIGO and VIRGO \cite{LIGOScientific:2016aoc}, several techniques of astronomical observations have been emerged. Observation of the first multi-messenger transient GW170817 \cite{LIGOScientific:2017vwq, LIGOScientific:2017ync} has raised interests to study the details of the physical processes in their sources from different perspectives. The gamma ray burst GRB 170817A was detected by the Fermi Gamma-ray Space Telescope and INTEGRAL 1.7 seconds after the gravitational wave signal GW170817, which was detected by the LIGO/Virgo collaboration in 2017. These signals were produced by the neutron star collision in the galaxy NGC 4993.
In the event of an electromagnetic counterpart coincident with a GW detection, direct constraints on $\Delta v_g$ can be placed based on the difference in arrival times between the coincident gravitational and electromagnetic signals as well as an estimate of the distance to the source. 
\subsubsection{Methods}
\label{sec:MM_constraint_methods}
Following the procedure of \cite{LIGOScientific:2017zic}, deviations from the classically predicted group velocity of GWs can be derived from measurements of the time delay between coincident GW and electromagnetic signals,
\begin{align}
    \frac{\Delta v}{v_{\text{EM}}} \approx v_{\text{EM}} \frac{\Delta t}{d_L},
    \label{eq:delta-vg-MM}
\end{align}
where $\Delta v = v_{\text{GW}} - v_{\text{EM}}$, $\Delta t$ is the time delay between the two signals, and $d_L$ is the luminosity distance to the source. The time delay that appears in Eq. \ref{eq:delta-vg-MM} is assumed to be caused purely from polymer effects. However, the observed time delay will in reality be a sum of the time delay due to polymer effects and any difference in the emission times of the GW and EM signals $\Delta t_{\mathrm{obs}} \equiv \Delta t_{\mathrm{poly}} + \Delta t_{\mathrm{lag}}$. While Ref. \cite{LIGOScientific:2017zic} predicts a 10 s lag time, others propose significantly longer lags \cite{Ciolfi_2015, 2015ApJ...802...95R} up to $\sim 1000$s. Initially, we take $\Delta t_{\mathrm{lag}}$ to be perfectly known, but later on we will explore the dependence of the polymer scale measurements on the choice of $\Delta t_{\mathrm{lag}}$.

We take both $d_L$ and $\Delta t_{\mathrm{obs}}$ to be random variates, where $p(d_L)$ is approximated from the publicly available posterior samples produced from LIGO parameter estimation analysis. The distribution $p(\Delta t)$ is instead assumed to be normally distributed with expectation value $E[\Delta t]=1.74$ s and standard deviation $\sigma = 0.05$ s, again in accordance with \cite{LIGOScientific:2017zic}. For their lower bound estimate, Ref. \cite{LIGOScientific:2017zic} assumes a $10$ s lag between the emission of the source's gravitational radiation and its associated GRB. We adopt an even more conservative lag of $3.48$ s, which is equivalent to simultaneous signal emission with polymer effects inducing a $1.74$ s lag in the GW arrival time over the distance traveled by both signals. Compact binary coalescences are expected to be strong GW radiators associated with a delayed emission of short gamma-ray up to a few seconds compared to the GW emission, given that the central engine is expected to form within a few seconds during the inspiral phase \cite{LIGOScientific:2012fcp, Finn:1999vh}. Therefore, an observer in direction of the outflow is expected to observe the GW/GRB signal with a delay up to a few seconds for the electromagnetic counterpart emission. Other models suggest significantly longer lags \cite{doi:10.1126/science.aap9811, 2015ApJ...802...95R}, but we take this a conservative estimate. \\
To compute the posterior distribution on $\Delta v_g$, we first build the probability distribution $p_{d_L}(d_L)$ for $d_L$ from a public library. We then construct a normal probability distribution $p(\Delta t)$ for time delay. The posterior distribution for $\Delta v_g$ is proportional to the product of marginalized posteriors $p_{d_L}(d_L)$ and $p(\Delta t)$, as mentioned earlier,
\begin{equation}
    p\left(\Delta v_g\right) = \int p_{\Delta t} (\Delta t) p_{d_L}(d_L)d(d_L).
\end{equation}
The constructed probability distribution on $\Delta t$ is related to the distribution of $p(\Delta v_g d_L)$, via a Jacobian transformation as \eqref{jacobi-trans} using relation \eqref{eq:delta-vg-MM}. The resulting distribution on $\Delta v$ can then be computed by integrating over the joint probability density function,
\begin{align}
    \label{p_vg_MM_integral}
    p\left(\Delta v_g\right) = \int_{-\infty}^{\infty} |d_L| p_{\Delta t}\left(\Delta v_g d_L \right)p_{d_L}(d_L)d(d_L),
\end{align}
where $v_{\text{EM}}$ has been set to $1$. Then the polymer scale distribution is then calculated from $p(\Delta v_g)$ according to section \ref{sec:poly-constraints}.
%
%
\subsubsection{Results}
\begin{figure}
    \centering
    \includegraphics[width=\textwidth]{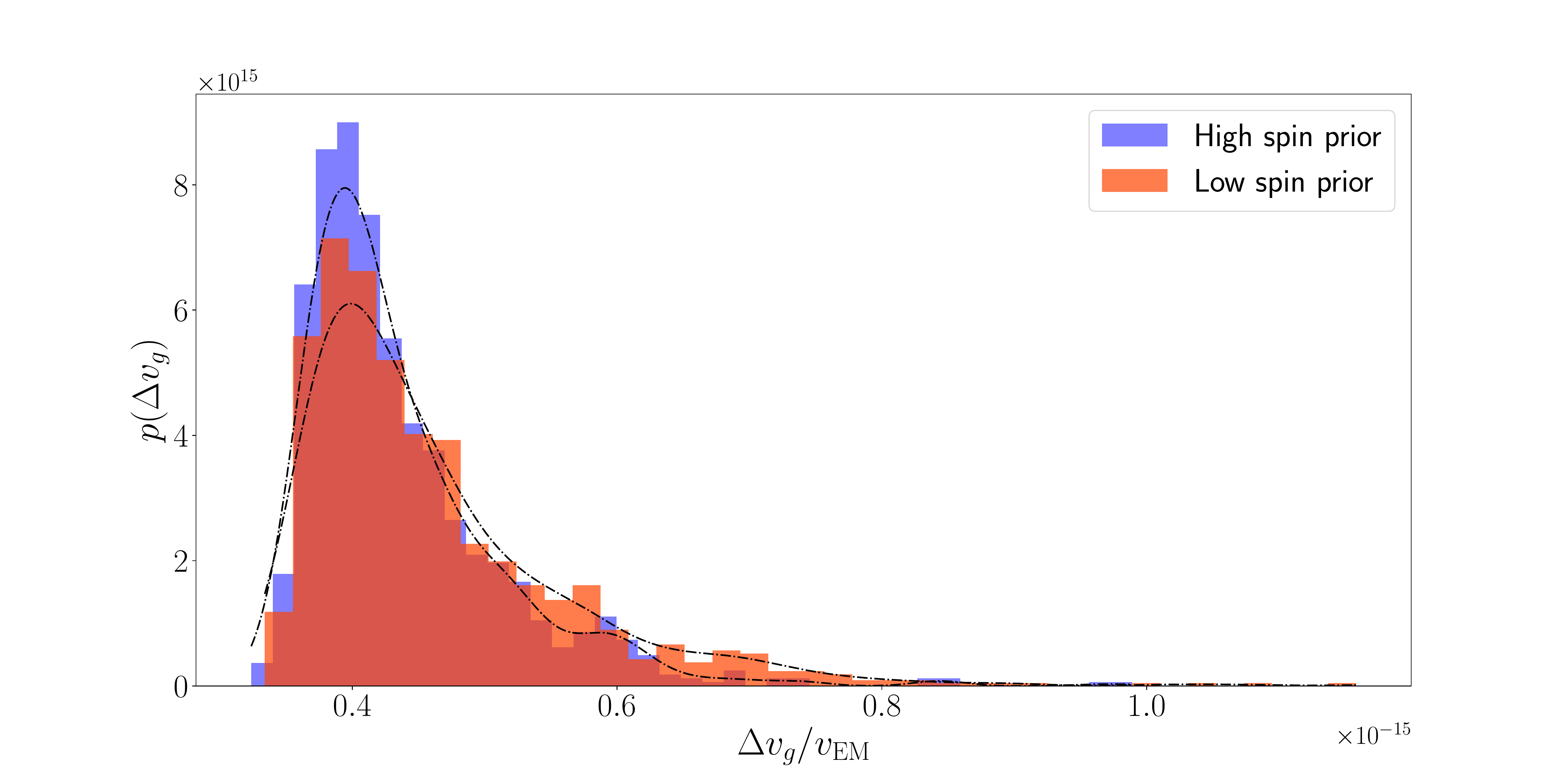}
    \caption{\small{Constraints on departure from classically predicted propagation speed of GWs, calculated based on estimates of luminosity distance to the source and time delay between GW170817 and GRB170817A. The maximum \textit{a posteriori} values for $\Delta v_g$ under the two spin priors are $\Delta v_{g,\text{low}}/v_{{\rm EM}} = 3.99 \protect\substack{+1.99 \\ -0.31} \times 10^{-16}$ and $\Delta v_{g, \mathrm{high}} /v_{{\rm EM}} = 3.94 \protect\substack{+1.44 \\ -0.31} \times 10^{-16} $.}}
    \label{fig:prob-vg-MM}
\end{figure}
The measured time delay between the arrival times of gravitational-wave detection GW170817 and the coincident gamma-ray detection GRB 170817A \cite{LIGOScientific:2017zic} constrains the deviations of the GW propagation speed from that of the electromagnetic radiation as,
\begin{align}\label{v-constraint}
    -3 \times 10^{-15} \leq \frac{\Delta v}{v_{\text{EM}}} \leq +7 \times 10^{-16},
\end{align}
assuming a conservative estimate of luminosity distance to the source binary of $26$ Mpc, the lower bound of the $90\%$ credible interval. The upper limit is unphysical when interpreting $\Delta v$ as a purely polymer-induced effect, so we restrict our estimate to the lower limit. This is equivalent to enforcing that the polymer scale must be real-valued. By inverting Eqs.~\eqref{rel:polyA-dev} and \eqref{rel:polyE-dev}, constraint \eqref{v-constraint} can be translated to the following upper bounds on the polymer corrections.
\begin{align}
U^{\mathcal{A}} \lesssim & 4 \times 5 \times 10^{-8}, \\
U^{\mathcal{E}}(k) \lesssim & \frac{4}{\sqrt{3}} \times 5 \times 10^{-8}.
\end{align}
What we have for polymer $\mathcal{E}$ is a frequency dependent correction, in line with the argument we presented and lead to equation \eqref{waveforms-2}, we replace its dependent with the max value, for the sake of simplicity and postpone its dependent to a future work.

Rewritting these expressions in terms of the $\bar{\mu}$ and $\bar{\nu}$ we obtain
\begin{align}
\bar{\mu} & = \frac{\hbar}{\bar{h} k} U^{\cal E}  \lesssim  10^{-7} \left( \frac{\hbar}{\bar{h} k} \right) \\
\bar{\nu} & = \frac{\hbar}{\bar{h} } U^{\cal A} \lesssim 10^{-7} \left( \frac{\hbar}{\bar{h} } \right)
\end{align}

\begin{figure}
    \centering
    \begin{subfigure}[b]{\textwidth}
    \includegraphics[width=\textwidth]{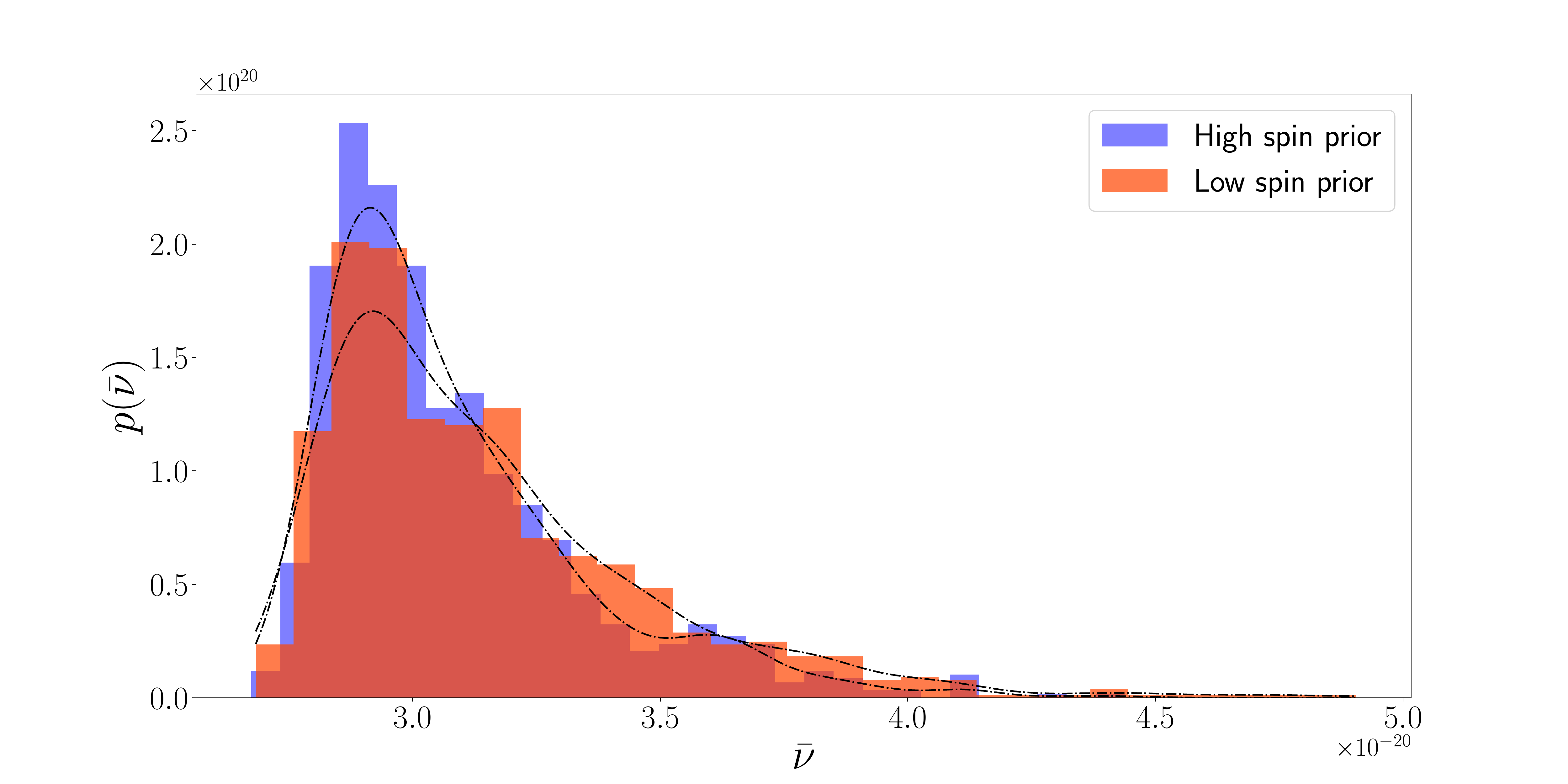}
    \caption{}
    \label{fig:polyA_MM_constraints}
    \end{subfigure}
    
    \begin{subfigure}[b]{\textwidth}
    \includegraphics[width=\textwidth]{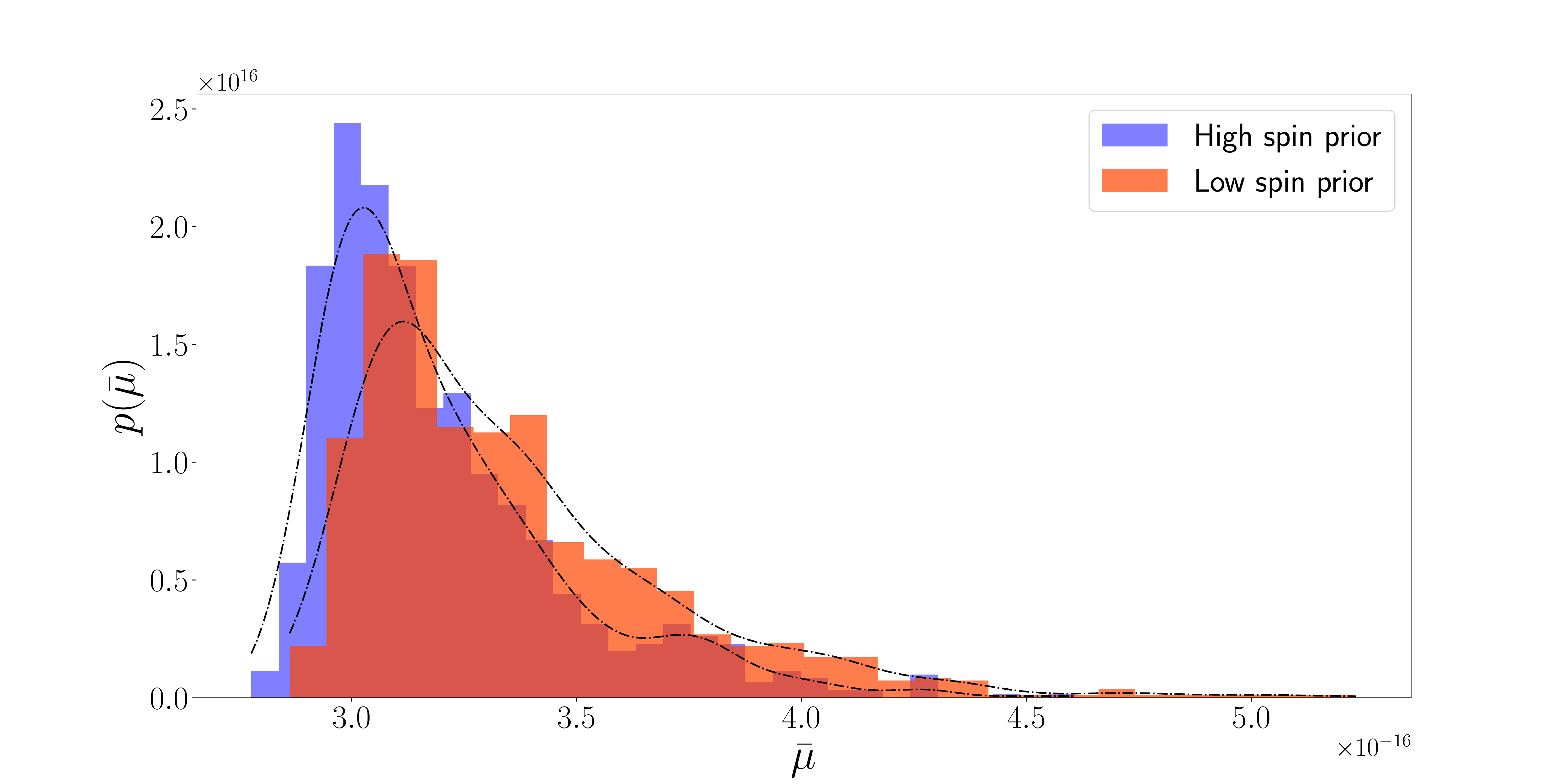}
    \caption{} \label{fig:polyE_MM_constraints}
    \end{subfigure}
\caption{\small{Constraints on polymer scales from coincident detections of GW170817 and GRB 170817A. The results for two priors on the neutron stars' spins are shown: one which disallows high spin values, and one which assumes all spin values are equally likely. The maximum \textit{a posteriori} values are $\bar{\nu}_{\mathrm{low}} = 2.91\protect\substack{+0.66 \\ -0.11}\times 10^{-20}, \; \bar{\nu}_{\mathrm{high}} = 2.91\protect\substack{+0.49 \\ -0.12}\times 10^{-20}$, $\bar{\mu}_{\mathrm{low}} =3.11\protect\substack{+0.70 \\ -0.12}\times 10^{-16}$ and $\bar{\mu}_{\mathrm{high}} = 3.03\protect\substack{+0.51 \\ -0.12}\times 10^{-16}$.}}
\label{fig:MM_constraints}
\end{figure}

In addition to this point-statistic bound, we also estimate the PDF on the luminosity distance to the source of GW170817 from posterior samples provided in the LVC's public data release\footnote{Official posterior samples for all source parameters, including $d_L$, can be found on the \href{https://www.gw-openscience.org/events/GW170817/}{GW170817 GWOSC page.}}. 
Approximating the PDF on the time delay as a Gaussian, we use Eq.~\eqref{p_vg_MM_integral} to compute the PDF on $\Delta v_g$, presented in Fig. \ref{fig:prob-vg-MM}. Finally, the resulting PDF on the polymer parameters under the two polymer quantization schemes are computed following the methods described in sections \ref{sec:poly-constraints} and \ref{sec:MM_constraint_methods}, depicted in Figs.~\ref{fig:MM_constraints}. We calculate the equivalent 90\% credible regions for the PDFs on the effective polymer parameters $\bar{\nu}$ and $\bar{\mu}$ and also bare parameters ${\nu}$ and ${\mu}$ report them in Table \ref{tab:MM_poly_table}.
\begin{table}[]
    \centering
    \begin{tabular}{|p{.3\textwidth}|p{.3\textwidth}|p{.3\textwidth}|}
    \hline
    \multicolumn{3}{|p{.9\textwidth}|}{\qquad \qquad \qquad \qquad Multi-messenger Constraints (GW170817 and GRB170817A)}\\
    \hline
    \qquad \qquad Observable/Spin  & \qquad \qquad Low spin prior & \qquad \qquad High spin prior \\
    \hline
     \qquad \qquad $\max(\Delta v_g)$ & \qquad \qquad $3.99\substack{+1.99 \\ -0.32}\times 10^{-16}$ & \qquad \qquad $3.94\substack{+1.44 \\ -0.31}\times 10^{-16}$ \\
    \hline
     \qquad \qquad $\bar{\nu}(\text{kg}\cdot \text{m}^2/\text{s})$ & \qquad \qquad $2.91\substack{+0.66 \\ -0.11}\times 10^{-20}$ & \qquad \qquad $2.91\substack{+0.49 \\ -0.12}\times 10^{-20}$\\
    \hline
    \qquad \qquad $\nu (\text{kg}^{1/2})$ & \qquad \qquad $2.66\substack{+0.60 \\ -0.10}\times 10^{-56}$ & \qquad \qquad $2.66\substack{+0.45 \\ -0.10}\times 10^{-56}$ \\
    \hline
    \qquad \qquad $\bar{\mu}(\text{kg} \cdot \text{m}^2) $ & \qquad \qquad $3.11\substack{+0.70 \\ -0.12}\times 10^{-16}$ & \qquad \qquad $3.03\substack{+0.51 \\ -0.12}\times 10^{-16}$\\
    \hline
    \qquad \qquad $\mu(\text{kg}^{1/2} \cdot \text{s})$ & \qquad \qquad $2.84\substack{+0.64 \\ -0.11}\times 10^{-52}$ & \qquad \qquad $2.76\substack{+0.46 \\ -0.11}\times 10^{-52}$ \\
    \hline
    \qquad GW strain at Peak & \qquad \qquad $2.88 \times 10^{-22}$ & \qquad \qquad $2.87 \times 10^{-22}$ \\
    \hline
    \qquad frequency at peak Hz & \qquad \qquad $2582.24$ & \qquad \qquad $2652.88$\\
    \hline
    \end{tabular}
    \caption{\small 90\% credible intervals for PDFs on $\{ \bar{\mu},\bar{\nu}\}_{\{\mathrm{low}, \mathrm{high}\}}$ shown in Fig.~\ref{fig:MM_constraints}.}
    \label{tab:MM_poly_table}
\end{table}
It should be noted that for this event, the choice of spin prior is particularly important. Higher spin values allow for the neutron star to sustain a higher mass, a parameter which is degenerate with luminosity distance at the level of the gravitational waveform. However, highly spinning neutron stars are thought to be rarer than ones with more moderate spins due to a loss of rotational energy through the powering of magnetically-driven plasma winds \cite{Goldreich:1969sb, Contopoulos_1999,Spitkovsky_2006}. Thus, Fig. \ref{fig:MM_constraints} includes two posterior density functions, corresponding to two choices of spin prior: one which restricts the spin parameters to low values, and one which assumes all values of the spin parameters are equally likely.
%
\begin{figure}
\centering
\begin{subfigure}[b]{\textwidth}
\includegraphics[width=.85\textwidth]{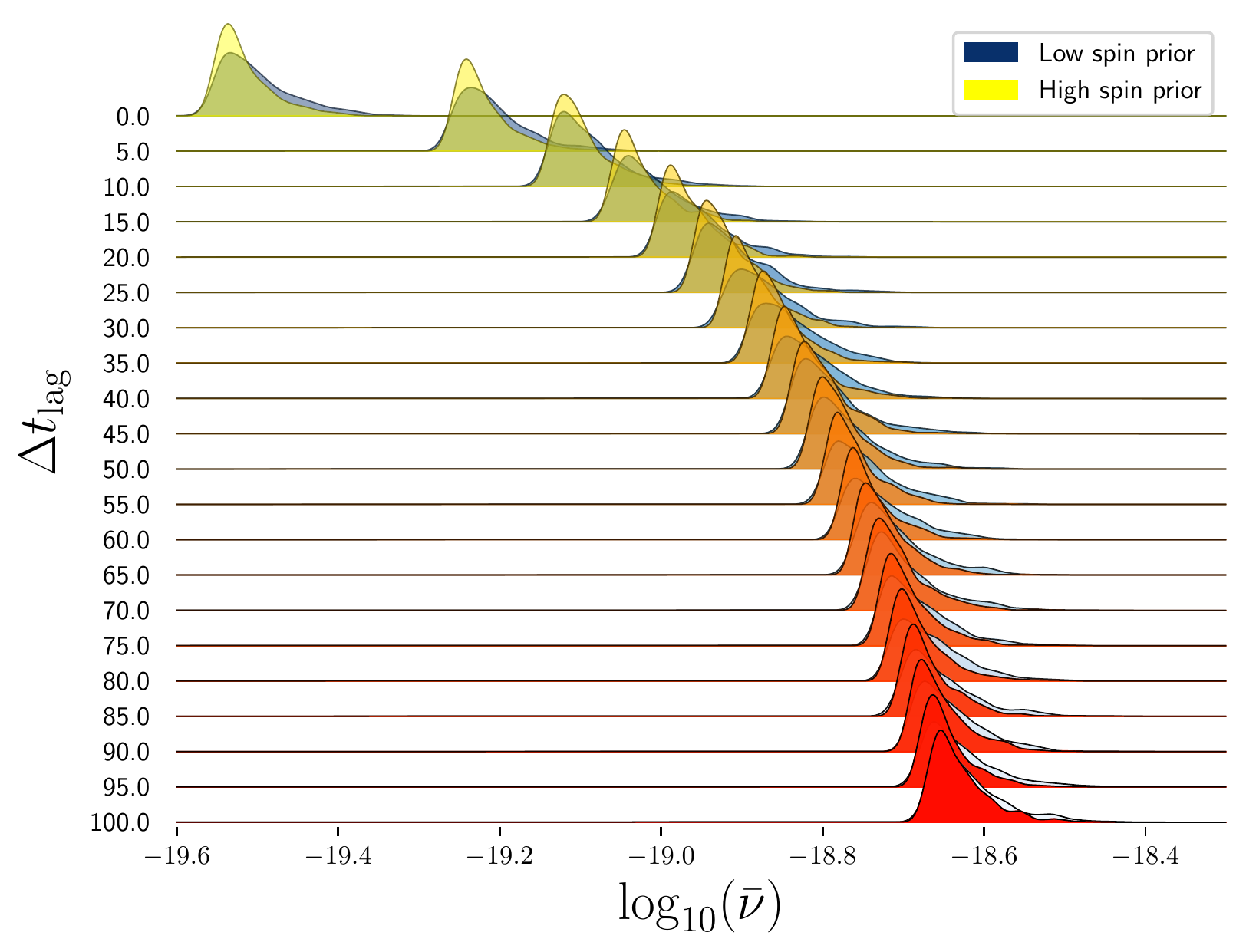}
\caption{Dependence of $p(\bar{\nu})$ on the lag between signal emission, $t_{0,\text{GW}} - t_{0, \text{EM}}$}
\label{fig:polyA_ridgeline}
\end{subfigure}
\begin{subfigure}[b]{\textwidth}
\includegraphics[width=.85\textwidth]{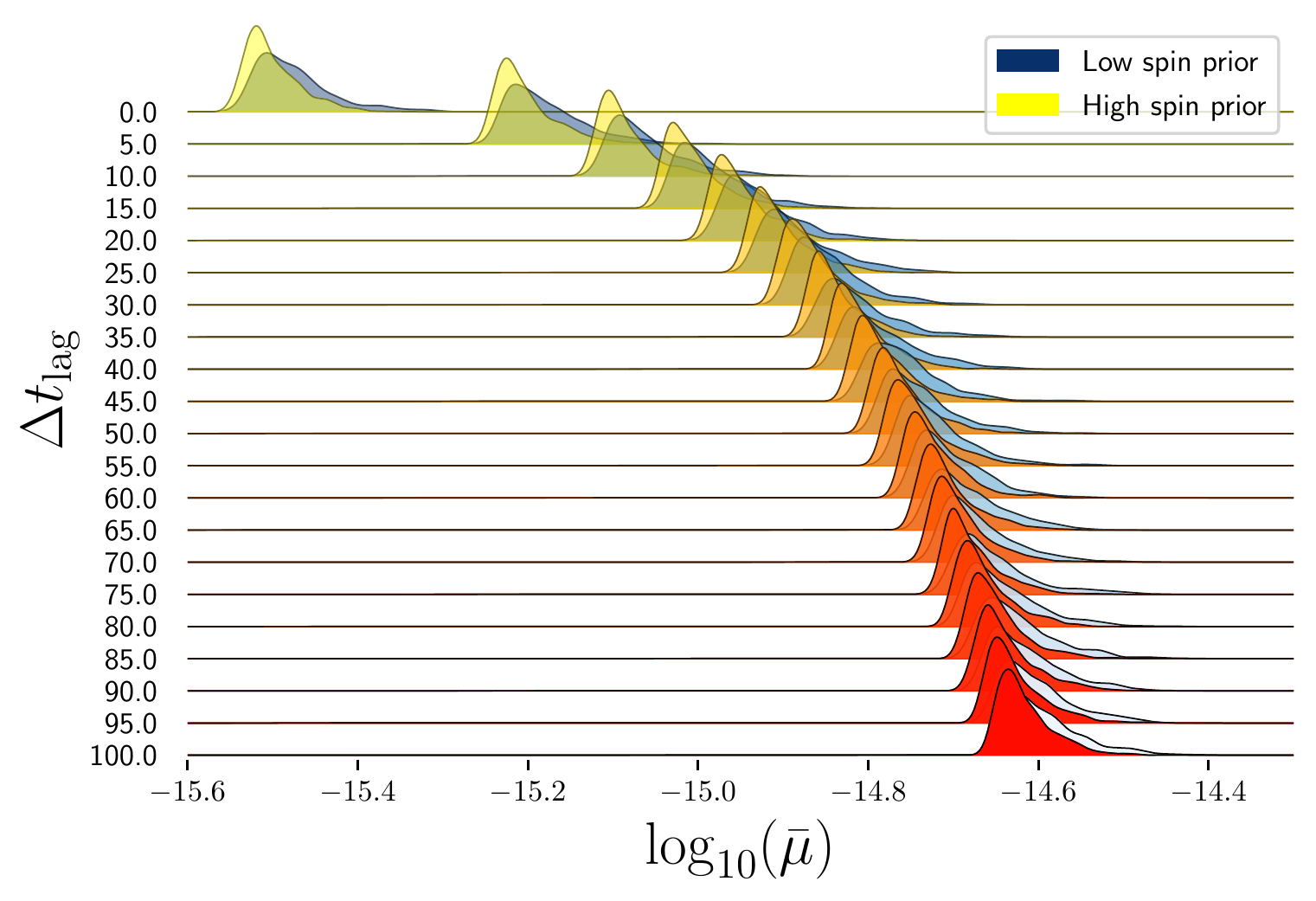}
\caption{Dependence of $p(\bar{\mu})$ on the lag between signal emission, $t_{0,\text{GW}} - t_{0, \text{EM}}$}
\label{fig:polyE_ridgeline}
\end{subfigure}
\caption{\small Dependence of constraints of polymer scales $\bar{\nu}$ and $\bar{\mu}$ on the time delay assumption for the GW and EM signals.}
\end{figure}
The dependence of polymer constraints on the assumption of time delay between GW signal emission and GRB emission is displayed in Figs.~\ref{fig:polyA_ridgeline} and \ref{fig:polyE_ridgeline} for both polymer quantization schemes and both choices for the spin prior on $d_L$. The means of the distributions on $\bar{\nu}$ are all within the range $-19.50 \leq \log_{10} \bar{\nu}_{\mathrm{low}} \leq -18.62$ and $-19.51 \leq \log_{10} \bar{\nu}_{\mathrm{high}} \leq -18.63$ for low and high values for spin prior. While the means on the distribution on $\bar{\mu}$ are all within the range $-15.47 \leq \log_{10} \bar{\mu}_{\mathrm{low}} \leq -14.59$ and $-15.50 \leq \log_{10} \bar{\mu}_{\mathrm{high}} \leq -14.62$ for low and high values. Additionally, all variances for the distributions on $\{ \bar{\mu},\bar{\nu}\}_{\{\mathrm{low}, \mathrm{high}\}} \leq 2.0 \times 10^{-3}$. These plots show that the estimated constraints does show strong dependence on the time delay assumption we made during our analysis.  


\section{Discussion and Conclusions\label{sec:conclusion}}

As presented in \cite{Garcia-Chung:2020zyq}, polymer quantization affects the wave-form and the propagation speed of gravitational waves and predicts departures from classical GR. To place constraint on the polymer scale, we use two procedures for leveraging polymer scale-dependent departures. In the first approach, we employed inter-detector arrival time differences for detected GW signals in observatories to find probability distribution function on deviation of the propagation speed of polymer GWs compared to their GR-predicted classical propagation speed, $\Delta v_g$. After simulating and extracting the strain and frequency of the signals at the peak, we translate the constraints from $\Delta v_g$ to polymer parameters $\nu$ and $\mu$. The details of posterior data for each signal can be found in Table~\ref{tab:peaks}. In the ``Combined(BBH)'' case, which we combined only the binary black hole events, we obtained constraints on polymer parameters, $ \nu = 0.96\substack{+0.15 \\ -0.21} \times 10^{-53} {\rm kg}^{1/2}$ and $\mu = 0.94\substack{+0.75 \\ -0.20} \times 10^{-48} {\rm kg}^{1/2} \cdot \text{s}$. After including the data for the single BNS event GW170817, the ``Combined'' value for $\nu$ and $\mu$ changed to $0.90\substack{+0.16 \\ -0.20} \times 10^{-53} {\rm kg}^{1/2}$ and $0.25\substack{+0.05 \\ -0.15} \times 10^{-48} {\rm kg}^{1/2} \cdot \text{s}$ respectively. The listed uncertainties are calculated to the $90\%$ credible level. In the second approach, we tried to find the constraints on the polymer parameters from a different method, by comparing the arrival time difference between multimessenger GW signal GW170817 with its EM counterpart GRB170817A. By assuming a conservative lag of 3.48 s, we extracted polymer scale constraints from $\max(\Delta v_g)$ for two spin priors, low and high. For the low spin case we find constrains $\nu = 2.66\substack{+0.60 \\ -0.10}\times 10^{-56} {\rm kg}^{1/2}$ and $\mu = 2.84\substack{+0.64 \\ -0.11}\times 10^{-52} {\rm kg}^{1/2}\cdot \text{s}$, respectively. For the high spin case the constraints turn out to be $\nu = 2.66\substack{+0.45 \\ -0.10}\times 10^{-56} {\rm kg}^{1/2}$ and $\mu = 2.76\substack{+0.46 \\ -0.11}\times 10^{-52} {\rm kg}^{1/2}\cdot \text{s}$. 

In our previous work \cite{Garcia-Chung:2022pdy}, from a completely different approach, we found bounds $ 10^{-52} < \nu < 10^{-58} $ and $ 10^{-44}< \mu < 10^{-50} $ for a given length scale $\ell$ for the binary system. Constraints on $\nu$ and $\mu$ extracted in the first procedure are within the detection range of LISA, while analysis in the second procedure shows that only quantum effects in the $\mathcal{A}$ scheme falls in the LISA range. One might tend to conclude that constrain on $\mu$ obtained from the second approach, makes predictions of the quantization scheme $\mathcal{E}$, undetectable in the future observation of LISA, but one should note that these effects are frequency-dependent, and thus by going to higher frequencies, their chance of detection increases. On the other hand, constraints reported in \cite{Garcia-Chung:2022pdy}, obtained for different length scale $\ell$ of the system, that means, we assumed different range of frequency for the GWs, by changing its value, different constraint can be extracted. It is not an odd feature, since, extracting any bounds for the polymer scale, is closely connected with the characteristic properties of the underling system, we will elaborate on this point when we compare our results with previously reported constraints. 

Although the main motivation of the polymer quantization comes from quantizing the gravitational degrees of freedom, but most of the previously reported constrains on the polymer scale obtained by considering matter fields in different setups \cite{Castellanos:2013ru, Chacon-Acosta:2014zva, Demir:2014rfa, Demarie:2012tz, Nozari:2015qoi, Khodadi:2017kah}. In almost all of them, the reported bound on the polymer scale was sensitive to the characteristic properties of the setup, for example in \cite{Castellanos:2013ru, Chacon-Acosta:2014zva}, by changing the number of particles and the characteristic length of the one–dimensional oscillator, a different bound on the polymer scale can be obtained, or in \cite{Demir:2014rfa, Demarie:2012tz}, different value for the number of particles, size of the system or barrier width, would result in different bound on the polymer parameter. Even in \cite{Bonder:2017ckx}, which employs the same procedure for the mode decomposition of the electromagnetic fields, final bounds on the polymer scale depends on the selected value for the size of the decomposition box and the amplitude of the observed GRB, which shows a similar role to the parameter $\ell$ in our setup. 
 
When constraints on $\Delta v_g$ are estimated from inter-detector time delays for pure GW signals, the resulting distributions on the polymer scales are relatively uninformative. The uncertainty on the distributions in Figs.~\ref{fig:polyA_pureGW_constraints} and \ref{fig:polyE_pureGW_constraints} are significantly larger than even theoretical constraints, with GW170817 providing the best single-event constraints due to its precise sky location measurement. There is, however, modest improvement when information from multiple GW events is combined. While our analysis only includes events from the LVC's first GW Transient Catalog (GWTC1) \cite{LIGOScientific:2018mvr}, the second \cite{LIGOScientific:2020ibl, LIGOScientific:2021usb} and third \cite{LIGOScientific:2021djp} catalogs add an additional 79 high-significance candidates to the list of GW detections. While additional constraints on the polymer scale from $\mathcal{O}(100)$ GW events may not be sufficient to make robust claims about the existence of polymer effects, next-generation GW detectors such as Cosmic Explorer (CE), and Einstein Telescope (ET) are set to provide $\mathcal{O}(10^6)$ observations by 2050 \cite{Iacovelli_2022}. Even in the likely case where polymer constraints do not improve with the current typical signal-to-noise ratio (SNR) of GW events (regardless of how many observations are made at that SNR), ET and CE are expected to make $\mathcal{O}(10^3)$ observations of BBH signals with SNR $\geq 100$.

GW events accompanied by an associated electromagnetic signal offer an opportunity to constrain deviations from the classically predicted GW propagation speed from time delays over astrophysical scales. This translates into many orders-of-magnitude tighter constraints on the polymer scale as shown in Fig.~\ref{fig:MM_constraints}. However, this approach is susceptible to systematics. Estimates of the time delay between the emission of GWs and GRBs in BNS systems varies widely which has a non-negligible impact on the resulting PDF on the polymer scales. We explore this dependency in Figs.~\ref{fig:polyA_ridgeline} and \ref{fig:polyE_ridgeline}. We find that the uncertainty is largely independent of the lag. While the mean of the PDF varies with the the emission time difference between the two signals, the means are within one order of magnitude of each other despite the emission delay ranging from 0 s to 100 s. This reflects the fact that when the propagation speed deviation is estimated as $\Delta v_g \approx \Delta t / d_L$, the polymer scales go as $U^{\{\mathcal{A}, \mathcal{E}\}} \sim \sqrt{\Delta t}$. With better BNS merger modeling in both the GW and EM sectors, this systematic dependence can be ameliorated yielding more trustworthy polymer constraints.

Furthermore, we note that Eq. \eqref{rel:polyE-dev} has a spectral dependence. Since GWs have a non-trivial frequency evolution (their amplitude and frequency change over the inspiral, merger and ringdown phases), we should consider $\Delta v^{\mathcal{E}} (k)$ as a function and $\mu$ as a free parameter of the model, and infer its value from the posterior analysis. However, since the posterior for $\Delta v_g$ was not binned in $k$-space (or equivalently, frequency space), we treated $\Delta v^{\mathcal{E}}$ as a free model parameter and mapped the posterior distribution from this parameter to the polymer parameter $\mu$. We also had to choose a constant value for the frequency and set $k=k_{\text{ value at the peak}}$, aiming for the most optimistic constraint for $\mu$. Frequency-dependent constraints are something we hope to explore in a future study, which would require having $v_g$ binned in the frequency space.

Our model shows that if spacetime is quantum with a minimal length scale, then this should result in the modification to the waveform of the gravitational waves, including their amplitude and dispersion relation, and in particular will lead to the dependence of the speed of propagation of gravitational waves on their frequency. If such dependence is actually experimentally established, then our model shows how to get an indirect bound on such minimum scale from the aforementioned dependence of the propagation speed on the frequency of the waves. This, together with our precise prediction to modification of the waveform can lead to two outcomes: either this precise waveform will match near-future precision experiments and the results match our predictions, in which case it would be an strong indication of the quantum nature of spacetime, or in case of disagreement with experiment, this specific polymer model will be refuted. In our opinion, either case would be fruitful results.  

Furthermore, although our results show that polymer effects will modify the propagation of GWs and an upper bound for the polymer parameters can be found using our suggested approach, nevertheless, in order to concretely obtain an indication of quantum gravity polymer effects, i.e., a smoking gun result, we need to also find lower bounds to the polymer parameters. A necessary (but probably not sufficient) improvement in this direction is to explore the aforementioned frequency dependency of the polymer parameters together with other wave-like effects, and particularly extending our model to cases where the background, as well as perturbations, is also quantum.

In future investigations we hope to perform a full forecasting study to  quantify the level at which additional GW events of a certain SNR improve polymer constraints estimated from inter-detector time delays. Additionally, with the first-order analytic approximations to the full polymer-corrected gravitational waveforms (Eqs. \eqref{eq:EoM-eff-A-sol-app2}, \eqref{eq:EoM-eff-E-sol-app2}), it is now feasible to directly constrain the polymer scale by performing Bayesian parameter estimation with waveforms that include polymer effects--a subject of another future study.

\begin{acknowledgments}

S.~R. acknowledges the support of the Natural Science and Engineering Research Council of Canada (NSERC) under funding reference numbers RGPIN-2021-03644 and DGECR-2021-00302. This research was also partially supported by the Perimeter Institute for Theoretical Physics, which is funded by the Government of Canada through the Department of Innovation, Science, and Economic Development and by the Province of Ontario through the Ministry of Research, Innovation and Science. 
Y.~T. expresses gratitude for the warm hospitality provided by the University of Warsaw where a portion of this work was completed.
S.~R., Y.~T. and A.~P conducted this work as part of the COST (European Cooperation in Science and Technology), Action CA18108: \emph{Quantum gravity phenomenology in the multi-messenger approach}.

\end{acknowledgments}



\bibliographystyle{apsrev4-2}
\bibliography{mainbib}

\begin{thebibliography}{47}%
\makeatletter
\providecommand \@ifxundefined [1]{%
 \@ifx{#1\undefined}
}%
\providecommand \@ifnum [1]{%
 \ifnum #1\expandafter \@firstoftwo
 \else \expandafter \@secondoftwo
 \fi
}%
\providecommand \@ifx [1]{%
 \ifx #1\expandafter \@firstoftwo
 \else \expandafter \@secondoftwo
 \fi
}%
\providecommand \natexlab [1]{#1}%
\providecommand \enquote  [1]{``#1''}%
\providecommand \bibnamefont  [1]{#1}%
\providecommand \bibfnamefont [1]{#1}%
\providecommand \citenamefont [1]{#1}%
\providecommand \href@noop [0]{\@secondoftwo}%
\providecommand \href [0]{\begingroup \@sanitize@url \@href}%
\providecommand \@href[1]{\@@startlink{#1}\@@href}%
\providecommand \@@href[1]{\endgroup#1\@@endlink}%
\providecommand \@sanitize@url [0]{\catcode `\\12\catcode `\$12\catcode
  `\&12\catcode `\#12\catcode `\^12\catcode `\_12\catcode `\%12\relax}%
\providecommand \@@startlink[1]{}%
\providecommand \@@endlink[0]{}%
\providecommand \url  [0]{\begingroup\@sanitize@url \@url }%
\providecommand \@url [1]{\endgroup\@href {#1}{\urlprefix }}%
\providecommand \urlprefix  [0]{URL }%
\providecommand \Eprint [0]{\href }%
\providecommand \doibase [0]{https://doi.org/}%
\providecommand \selectlanguage [0]{\@gobble}%
\providecommand \bibinfo  [0]{\@secondoftwo}%
\providecommand \bibfield  [0]{\@secondoftwo}%
\providecommand \translation [1]{[#1]}%
\providecommand \BibitemOpen [0]{}%
\providecommand \bibitemStop [0]{}%
\providecommand \bibitemNoStop [0]{.\EOS\space}%
\providecommand \EOS [0]{\spacefactor3000\relax}%
\providecommand \BibitemShut  [1]{\csname bibitem#1\endcsname}%
\let\auto@bib@innerbib\@empty
\bibitem [{\citenamefont {Arun}\ \emph {et~al.}(2022)\citenamefont {Arun} \emph
  {et~al.}}]{LISA:2022kgy}%
  \BibitemOpen
  \bibfield  {author} {\bibinfo {author} {\bibfnamefont {K.~G.}\ \bibnamefont
  {Arun}} \emph {et~al.} (\bibinfo {collaboration} {LISA}),\ }\href
  {https://doi.org/10.1007/s41114-022-00036-9} {\bibfield  {journal} {\bibinfo
  {journal} {Living Rev. Rel.}\ }\textbf {\bibinfo {volume} {25}},\ \bibinfo
  {pages} {4} (\bibinfo {year} {2022})},\ \Eprint
  {https://arxiv.org/abs/2205.01597} {arXiv:2205.01597 [gr-qc]} \BibitemShut
  {NoStop}%
\bibitem [{\citenamefont {Auclair}\ \emph {et~al.}(2023)\citenamefont {Auclair}
  \emph {et~al.}}]{LISACosmologyWorkingGroup:2022jok}%
  \BibitemOpen
  \bibfield  {author} {\bibinfo {author} {\bibfnamefont {P.}~\bibnamefont
  {Auclair}} \emph {et~al.} (\bibinfo {collaboration} {LISA Cosmology Working
  Group}),\ }\href {https://doi.org/10.1007/s41114-023-00045-2} {\bibfield
  {journal} {\bibinfo  {journal} {Living Rev. Rel.}\ }\textbf {\bibinfo
  {volume} {26}},\ \bibinfo {pages} {5} (\bibinfo {year} {2023})},\ \Eprint
  {https://arxiv.org/abs/2204.05434} {arXiv:2204.05434 [astro-ph.CO]}
  \BibitemShut {NoStop}%
\bibitem [{\citenamefont {Addazi}\ \emph {et~al.}(2022)\citenamefont {Addazi}
  \emph {et~al.}}]{Addazi:2021xuf}%
  \BibitemOpen
  \bibfield  {author} {\bibinfo {author} {\bibfnamefont {A.}~\bibnamefont
  {Addazi}} \emph {et~al.},\ }\href
  {https://doi.org/10.1016/j.ppnp.2022.103948} {\bibfield  {journal} {\bibinfo
  {journal} {Prog. Part. Nucl. Phys.}\ }\textbf {\bibinfo {volume} {125}},\
  \bibinfo {pages} {103948} (\bibinfo {year} {2022})},\ \Eprint
  {https://arxiv.org/abs/2111.05659} {arXiv:2111.05659 [hep-ph]} \BibitemShut
  {NoStop}%
\bibitem [{\citenamefont {Thiemann}(2007)}]{Thiemann:2007pyv}%
  \BibitemOpen
  \bibfield  {author} {\bibinfo {author} {\bibfnamefont {T.}~\bibnamefont
  {Thiemann}},\ }\href {https://doi.org/10.1017/CBO9780511755682} {\emph
  {\bibinfo {title} {{Modern Canonical Quantum General Relativity}}}},\
  Cambridge Monographs on Mathematical Physics\ (\bibinfo  {publisher}
  {Cambridge University Press},\ \bibinfo {year} {2007})\BibitemShut {NoStop}%
\bibitem [{\citenamefont {Rovelli}(2004)}]{Rovelli:2004tv}%
  \BibitemOpen
  \bibfield  {author} {\bibinfo {author} {\bibfnamefont {C.}~\bibnamefont
  {Rovelli}},\ }\href {https://doi.org/10.1017/CBO9780511755804} {\emph
  {\bibinfo {title} {{Quantum gravity}}}},\ Cambridge Monographs on
  Mathematical Physics\ (\bibinfo  {publisher} {Univ. Pr.},\ \bibinfo {address}
  {Cambridge, UK},\ \bibinfo {year} {2004})\BibitemShut {NoStop}%
\bibitem [{\citenamefont {Gambini}\ and\ \citenamefont
  {Pullin}(2011)}]{Gambini:2011zz}%
  \BibitemOpen
  \bibfield  {author} {\bibinfo {author} {\bibfnamefont {R.}~\bibnamefont
  {Gambini}}\ and\ \bibinfo {author} {\bibfnamefont {J.}~\bibnamefont
  {Pullin}},\ }\href
  {https://doi.org/10.1093/acprof:oso/9780199590759.001.0001} {\emph {\bibinfo
  {title} {{A First Course in Loop Quantum Gravity}}}}\ (\bibinfo  {publisher}
  {Oxford University Press},\ \bibinfo {year} {2011})\BibitemShut {NoStop}%
\bibitem [{\citenamefont {Ashtekar}\ \emph {et~al.}(2003)\citenamefont
  {Ashtekar}, \citenamefont {Fairhurst},\ and\ \citenamefont
  {Willis}}]{Ashtekar:2002sn}%
  \BibitemOpen
  \bibfield  {author} {\bibinfo {author} {\bibfnamefont {A.}~\bibnamefont
  {Ashtekar}}, \bibinfo {author} {\bibfnamefont {S.}~\bibnamefont
  {Fairhurst}},\ and\ \bibinfo {author} {\bibfnamefont {J.~L.}\ \bibnamefont
  {Willis}},\ }\href {https://doi.org/10.1088/0264-9381/20/6/302} {\bibfield
  {journal} {\bibinfo  {journal} {Class. Quant. Grav.}\ }\textbf {\bibinfo
  {volume} {20}},\ \bibinfo {pages} {1031} (\bibinfo {year} {2003})},\ \Eprint
  {https://arxiv.org/abs/gr-qc/0207106} {arXiv:gr-qc/0207106} \BibitemShut
  {NoStop}%
\bibitem [{\citenamefont {Morales-T\'{e}cotl}\ \emph
  {et~al.}(2015)\citenamefont {Morales-T\'{e}cotl}, \citenamefont
  {Orozco-Borunda},\ and\ \citenamefont {Rastgoo}}]{Tecotl:2015cya}%
  \BibitemOpen
  \bibfield  {author} {\bibinfo {author} {\bibfnamefont {H.~A.}\ \bibnamefont
  {Morales-T\'{e}cotl}}, \bibinfo {author} {\bibfnamefont {D.~H.}\ \bibnamefont
  {Orozco-Borunda}},\ and\ \bibinfo {author} {\bibfnamefont {S.}~\bibnamefont
  {Rastgoo}},\ }\href {https://doi.org/10.1103/PhysRevD.92.104029} {\bibfield
  {journal} {\bibinfo  {journal} {Phys.Rev.D}\ }\textbf {\bibinfo {volume}
  {92}},\ \bibinfo {pages} {104029} (\bibinfo {year} {2015})},\ \Eprint
  {https://arxiv.org/abs/1507.08651} {arXiv:1507.08651 [gr-qc]} \BibitemShut
  {NoStop}%
\bibitem [{\citenamefont {Morales-T\'{e}cotl}\ \emph
  {et~al.}(2017)\citenamefont {Morales-T\'{e}cotl}, \citenamefont {Rastgoo},\
  and\ \citenamefont {Ruelas}}]{Morales-Tecotl:2016ijb}%
  \BibitemOpen
  \bibfield  {author} {\bibinfo {author} {\bibfnamefont {H.~A.}\ \bibnamefont
  {Morales-T\'{e}cotl}}, \bibinfo {author} {\bibfnamefont {S.}~\bibnamefont
  {Rastgoo}},\ and\ \bibinfo {author} {\bibfnamefont {J.~C.}\ \bibnamefont
  {Ruelas}},\ }\href {https://doi.org/10.1103/PhysRevD.95.065026} {\bibfield
  {journal} {\bibinfo  {journal} {Phys.Rev.D}\ }\textbf {\bibinfo {volume}
  {95}},\ \bibinfo {pages} {065026} (\bibinfo {year} {2017})},\ \Eprint
  {https://arxiv.org/abs/1608.04498} {arXiv:1608.04498 [gr-qc]} \BibitemShut
  {NoStop}%
\bibitem [{\citenamefont {Bonder}\ \emph {et~al.}(2017)\citenamefont {Bonder},
  \citenamefont {Garcia-Chung},\ and\ \citenamefont
  {Rastgoo}}]{Bonder:2017ckx}%
  \BibitemOpen
  \bibfield  {author} {\bibinfo {author} {\bibfnamefont {Y.}~\bibnamefont
  {Bonder}}, \bibinfo {author} {\bibfnamefont {A.}~\bibnamefont
  {Garcia-Chung}},\ and\ \bibinfo {author} {\bibfnamefont {S.}~\bibnamefont
  {Rastgoo}},\ }\href {https://doi.org/10.1103/PhysRevD.96.106021} {\bibfield
  {journal} {\bibinfo  {journal} {Phys.Rev.D}\ }\textbf {\bibinfo {volume}
  {96}},\ \bibinfo {pages} {106021} (\bibinfo {year} {2017})},\ \Eprint
  {https://arxiv.org/abs/1704.08750} {arXiv:1704.08750 [gr-qc]} \BibitemShut
  {NoStop}%
\bibitem [{\citenamefont {Garcia-Chung}\ \emph
  {et~al.}(2021{\natexlab{a}})\citenamefont {Garcia-Chung}, \citenamefont
  {Mertens}, \citenamefont {Rastgoo}, \citenamefont {Tavakoli},\ and\
  \citenamefont {Vargas~Moniz}}]{Garcia-Chung:2020zyq}%
  \BibitemOpen
  \bibfield  {author} {\bibinfo {author} {\bibfnamefont {A.}~\bibnamefont
  {Garcia-Chung}}, \bibinfo {author} {\bibfnamefont {J.~B.}\ \bibnamefont
  {Mertens}}, \bibinfo {author} {\bibfnamefont {S.}~\bibnamefont {Rastgoo}},
  \bibinfo {author} {\bibfnamefont {Y.}~\bibnamefont {Tavakoli}},\ and\
  \bibinfo {author} {\bibfnamefont {P.}~\bibnamefont {Vargas~Moniz}},\ }\href
  {https://doi.org/10.1103/PhysRevD.103.084053} {\bibfield  {journal} {\bibinfo
   {journal} {Phys. Rev. D}\ }\textbf {\bibinfo {volume} {103}},\ \bibinfo
  {pages} {084053} (\bibinfo {year} {2021}{\natexlab{a}})},\ \Eprint
  {https://arxiv.org/abs/2012.09366} {arXiv:2012.09366 [gr-qc]} \BibitemShut
  {NoStop}%
\bibitem [{\citenamefont {Garcia-Chung}\ \emph {et~al.}(2022)\citenamefont
  {Garcia-Chung}, \citenamefont {Carney}, \citenamefont {Mertens},
  \citenamefont {Parvizi}, \citenamefont {Rastgoo},\ and\ \citenamefont
  {Tavakoli}}]{Garcia-Chung:2022pdy}%
  \BibitemOpen
  \bibfield  {author} {\bibinfo {author} {\bibfnamefont {A.}~\bibnamefont
  {Garcia-Chung}}, \bibinfo {author} {\bibfnamefont {M.~F.}\ \bibnamefont
  {Carney}}, \bibinfo {author} {\bibfnamefont {J.~B.}\ \bibnamefont {Mertens}},
  \bibinfo {author} {\bibfnamefont {A.}~\bibnamefont {Parvizi}}, \bibinfo
  {author} {\bibfnamefont {S.}~\bibnamefont {Rastgoo}},\ and\ \bibinfo {author}
  {\bibfnamefont {Y.}~\bibnamefont {Tavakoli}},\ }\href
  {https://doi.org/10.1088/1475-7516/2022/11/054} {\bibfield  {journal}
  {\bibinfo  {journal} {JCAP}\ }\textbf {\bibinfo {volume} {11}},\ \bibinfo
  {pages} {054}},\ \Eprint {https://arxiv.org/abs/2208.09739} {arXiv:2208.09739
  [gr-qc]} \BibitemShut {NoStop}%
\bibitem [{\citenamefont {Garcia-Chung}\ \emph
  {et~al.}(2021{\natexlab{b}})\citenamefont {Garcia-Chung}, \citenamefont
  {Mertens}, \citenamefont {Rastgoo}, \citenamefont {Tavakoli},\ and\
  \citenamefont {Moniz}}]{Garcia-Chung:2021doi}%
  \BibitemOpen
  \bibfield  {author} {\bibinfo {author} {\bibfnamefont {A.}~\bibnamefont
  {Garcia-Chung}}, \bibinfo {author} {\bibfnamefont {J.~B.}\ \bibnamefont
  {Mertens}}, \bibinfo {author} {\bibfnamefont {S.}~\bibnamefont {Rastgoo}},
  \bibinfo {author} {\bibfnamefont {Y.}~\bibnamefont {Tavakoli}},\ and\
  \bibinfo {author} {\bibfnamefont {P.~V.}\ \bibnamefont {Moniz}},\ }in\ \href
  {https://doi.org/10.1142/9789811269776_0347} {\emph {\bibinfo {booktitle}
  {{16th Marcel Grossmann Meeting on~Recent Developments in Theoretical and
  Experimental General Relativity, Astrophysics and Relativistic Field
  Theories}}}}\ (\bibinfo {year} {2021})\ \Eprint
  {https://arxiv.org/abs/2111.00292} {arXiv:2111.00292 [gr-qc]} \BibitemShut
  {NoStop}%
\bibitem [{\citenamefont {Cardoso}\ \emph
  {et~al.}(2016{\natexlab{a}})\citenamefont {Cardoso}, \citenamefont {Hopper},
  \citenamefont {Macedo}, \citenamefont {Palenzuela},\ and\ \citenamefont
  {Pani}}]{Cardoso:2016oxy}%
  \BibitemOpen
  \bibfield  {author} {\bibinfo {author} {\bibfnamefont {V.}~\bibnamefont
  {Cardoso}}, \bibinfo {author} {\bibfnamefont {S.}~\bibnamefont {Hopper}},
  \bibinfo {author} {\bibfnamefont {C.~F.~B.}\ \bibnamefont {Macedo}}, \bibinfo
  {author} {\bibfnamefont {C.}~\bibnamefont {Palenzuela}},\ and\ \bibinfo
  {author} {\bibfnamefont {P.}~\bibnamefont {Pani}},\ }\href
  {https://doi.org/10.1103/PhysRevD.94.084031} {\bibfield  {journal} {\bibinfo
  {journal} {Phys. Rev. D}\ }\textbf {\bibinfo {volume} {94}},\ \bibinfo
  {pages} {084031} (\bibinfo {year} {2016}{\natexlab{a}})},\ \Eprint
  {https://arxiv.org/abs/1608.08637} {arXiv:1608.08637 [gr-qc]} \BibitemShut
  {NoStop}%
\bibitem [{\citenamefont {Cardoso}\ \emph
  {et~al.}(2016{\natexlab{b}})\citenamefont {Cardoso}, \citenamefont
  {Franzin},\ and\ \citenamefont {Pani}}]{Cardoso:2016rao}%
  \BibitemOpen
  \bibfield  {author} {\bibinfo {author} {\bibfnamefont {V.}~\bibnamefont
  {Cardoso}}, \bibinfo {author} {\bibfnamefont {E.}~\bibnamefont {Franzin}},\
  and\ \bibinfo {author} {\bibfnamefont {P.}~\bibnamefont {Pani}},\ }\href
  {https://doi.org/10.1103/PhysRevLett.116.171101} {\bibfield  {journal}
  {\bibinfo  {journal} {Phys. Rev. Lett.}\ }\textbf {\bibinfo {volume} {116}},\
  \bibinfo {pages} {171101} (\bibinfo {year} {2016}{\natexlab{b}})},\ \bibinfo
  {note} {[Erratum: Phys.Rev.Lett. 117, 089902 (2016)]},\ \Eprint
  {https://arxiv.org/abs/1602.07309} {arXiv:1602.07309 [gr-qc]} \BibitemShut
  {NoStop}%
\bibitem [{\citenamefont {Abedi}\ \emph {et~al.}(2017)\citenamefont {Abedi},
  \citenamefont {Dykaar},\ and\ \citenamefont {Afshordi}}]{Abedi:2016hgu}%
  \BibitemOpen
  \bibfield  {author} {\bibinfo {author} {\bibfnamefont {J.}~\bibnamefont
  {Abedi}}, \bibinfo {author} {\bibfnamefont {H.}~\bibnamefont {Dykaar}},\ and\
  \bibinfo {author} {\bibfnamefont {N.}~\bibnamefont {Afshordi}},\ }\href
  {https://doi.org/10.1103/PhysRevD.96.082004} {\bibfield  {journal} {\bibinfo
  {journal} {Phys. Rev. D}\ }\textbf {\bibinfo {volume} {96}},\ \bibinfo
  {pages} {082004} (\bibinfo {year} {2017})},\ \Eprint
  {https://arxiv.org/abs/1612.00266} {arXiv:1612.00266 [gr-qc]} \BibitemShut
  {NoStop}%
\bibitem [{\citenamefont {Barcel\'o}\ \emph {et~al.}(2017)\citenamefont
  {Barcel\'o}, \citenamefont {Carballo-Rubio},\ and\ \citenamefont
  {Garay}}]{Barcelo:2017lnx}%
  \BibitemOpen
  \bibfield  {author} {\bibinfo {author} {\bibfnamefont {C.}~\bibnamefont
  {Barcel\'o}}, \bibinfo {author} {\bibfnamefont {R.}~\bibnamefont
  {Carballo-Rubio}},\ and\ \bibinfo {author} {\bibfnamefont {L.~J.}\
  \bibnamefont {Garay}},\ }\href {https://doi.org/10.1007/JHEP05(2017)054}
  {\bibfield  {journal} {\bibinfo  {journal} {JHEP}\ }\textbf {\bibinfo
  {volume} {05}},\ \bibinfo {pages} {054}},\ \Eprint
  {https://arxiv.org/abs/1701.09156} {arXiv:1701.09156 [gr-qc]} \BibitemShut
  {NoStop}%
\bibitem [{\citenamefont {Amelino-Camelia}\ \emph {et~al.}(1998)\citenamefont
  {Amelino-Camelia}, \citenamefont {Ellis}, \citenamefont {Mavromatos},
  \citenamefont {Nanopoulos},\ and\ \citenamefont
  {Sarkar}}]{AmelinoCamelia:1997gz}%
  \BibitemOpen
  \bibfield  {author} {\bibinfo {author} {\bibfnamefont {G.}~\bibnamefont
  {Amelino-Camelia}}, \bibinfo {author} {\bibfnamefont {J.~R.}\ \bibnamefont
  {Ellis}}, \bibinfo {author} {\bibfnamefont {N.~E.}\ \bibnamefont
  {Mavromatos}}, \bibinfo {author} {\bibfnamefont {D.~V.}\ \bibnamefont
  {Nanopoulos}},\ and\ \bibinfo {author} {\bibfnamefont {S.}~\bibnamefont
  {Sarkar}},\ }\href {https://doi.org/10.1038/31647} {\bibfield  {journal}
  {\bibinfo  {journal} {Nature}\ }\textbf {\bibinfo {volume} {393}},\ \bibinfo
  {pages} {763} (\bibinfo {year} {1998})},\ \Eprint
  {https://arxiv.org/abs/astro-ph/9712103} {arXiv:astro-ph/9712103 [astro-ph]}
  \BibitemShut {NoStop}%
\bibitem [{\citenamefont {Abbott}\ \emph
  {et~al.}(2017{\natexlab{a}})\citenamefont {Abbott} \emph
  {et~al.}}]{LIGOScientific:2017vwq}%
  \BibitemOpen
  \bibfield  {author} {\bibinfo {author} {\bibfnamefont {B.~P.}\ \bibnamefont
  {Abbott}} \emph {et~al.} (\bibinfo {collaboration} {LIGO Scientific,
  Virgo}),\ }\href {https://doi.org/10.1103/PhysRevLett.119.161101} {\bibfield
  {journal} {\bibinfo  {journal} {Phys. Rev. Lett.}\ }\textbf {\bibinfo
  {volume} {119}},\ \bibinfo {pages} {161101} (\bibinfo {year}
  {2017}{\natexlab{a}})},\ \Eprint {https://arxiv.org/abs/1710.05832}
  {arXiv:1710.05832 [gr-qc]} \BibitemShut {NoStop}%
\bibitem [{\citenamefont {Abbott}\ \emph
  {et~al.}(2017{\natexlab{b}})\citenamefont {Abbott} \emph
  {et~al.}}]{LIGOScientific:2017zic}%
  \BibitemOpen
  \bibfield  {author} {\bibinfo {author} {\bibfnamefont {B.~P.}\ \bibnamefont
  {Abbott}} \emph {et~al.} (\bibinfo {collaboration} {LIGO Scientific, Virgo,
  Fermi-GBM, INTEGRAL}),\ }\href {https://doi.org/10.3847/2041-8213/aa920c}
  {\bibfield  {journal} {\bibinfo  {journal} {Astrophys. J. Lett.}\ }\textbf
  {\bibinfo {volume} {848}},\ \bibinfo {pages} {L13} (\bibinfo {year}
  {2017}{\natexlab{b}})},\ \Eprint {https://arxiv.org/abs/1710.05834}
  {arXiv:1710.05834 [astro-ph.HE]} \BibitemShut {NoStop}%
\bibitem [{\citenamefont {Abbott}\ \emph {et~al.}(2019)\citenamefont {Abbott}
  \emph {et~al.}}]{LIGOScientific:2018mvr}%
  \BibitemOpen
  \bibfield  {author} {\bibinfo {author} {\bibfnamefont {B.~P.}\ \bibnamefont
  {Abbott}} \emph {et~al.} (\bibinfo {collaboration} {LIGO Scientific,
  Virgo}),\ }\href {https://doi.org/10.1103/PhysRevX.9.031040} {\bibfield
  {journal} {\bibinfo  {journal} {Phys. Rev. X}\ }\textbf {\bibinfo {volume}
  {9}},\ \bibinfo {pages} {031040} (\bibinfo {year} {2019})},\ \Eprint
  {https://arxiv.org/abs/1811.12907} {arXiv:1811.12907 [astro-ph.HE]}
  \BibitemShut {NoStop}%
\bibitem [{\citenamefont {Austrich-Olivares}\ \emph {et~al.}(2017)\citenamefont
  {Austrich-Olivares}, \citenamefont {Garcia-Chung},\ and\ \citenamefont
  {Vergara}}]{austrich2017instanton}%
  \BibitemOpen
  \bibfield  {author} {\bibinfo {author} {\bibfnamefont {J.~A.}\ \bibnamefont
  {Austrich-Olivares}}, \bibinfo {author} {\bibfnamefont {A.}~\bibnamefont
  {Garcia-Chung}},\ and\ \bibinfo {author} {\bibfnamefont {J.~D.}\ \bibnamefont
  {Vergara}},\ }\href {https://doi.org/10.1088/1361-6382/aa6d88} {\bibfield
  {journal} {\bibinfo  {journal} {Classical and Quantum Gravity}\ }\textbf
  {\bibinfo {volume} {34}},\ \bibinfo {pages} {115005} (\bibinfo {year}
  {2017})},\ \Eprint {https://arxiv.org/abs/hep-th/1604.07288}
  {arXiv:hep-th/1604.07288 [hep-th]} \BibitemShut {NoStop}%
\bibitem [{\citenamefont {Parra}\ and\ \citenamefont
  {Vergara}(2014)}]{parra2014polymer}%
  \BibitemOpen
  \bibfield  {author} {\bibinfo {author} {\bibfnamefont {L.}~\bibnamefont
  {Parra}}\ and\ \bibinfo {author} {\bibfnamefont {J.~D.}\ \bibnamefont
  {Vergara}},\ }in\ \href@noop {} {\emph {\bibinfo {booktitle} {AIP Conference
  Proceedings}}},\ Vol.\ \bibinfo {volume} {1577}\ (\bibinfo {organization}
  {American Institute of Physics},\ \bibinfo {year} {2014})\ pp.\ \bibinfo
  {pages} {269--280}\BibitemShut {NoStop}%
\bibitem [{\citenamefont {Liu}\ \emph {et~al.}(2020)\citenamefont {Liu},
  \citenamefont {He}, \citenamefont {Mikulski}, \citenamefont {Palenova},
  \citenamefont {Williams}, \citenamefont {Creighton},\ and\ \citenamefont
  {Tasson}}]{Liu:2020}%
  \BibitemOpen
  \bibfield  {author} {\bibinfo {author} {\bibfnamefont {X.}~\bibnamefont
  {Liu}}, \bibinfo {author} {\bibfnamefont {V.~F.}\ \bibnamefont {He}},
  \bibinfo {author} {\bibfnamefont {T.~M.}\ \bibnamefont {Mikulski}}, \bibinfo
  {author} {\bibfnamefont {D.}~\bibnamefont {Palenova}}, \bibinfo {author}
  {\bibfnamefont {C.~E.}\ \bibnamefont {Williams}}, \bibinfo {author}
  {\bibfnamefont {J.}~\bibnamefont {Creighton}},\ and\ \bibinfo {author}
  {\bibfnamefont {J.~D.}\ \bibnamefont {Tasson}},\ }\href
  {https://doi.org/10.1103/PhysRevD.102.024028} {\bibfield  {journal} {\bibinfo
   {journal} {Phys. Rev. D}\ }\textbf {\bibinfo {volume} {102}},\ \bibinfo
  {pages} {024028} (\bibinfo {year} {2020})}\BibitemShut {NoStop}%
\bibitem [{\citenamefont {Cornish}\ \emph {et~al.}(2017)\citenamefont
  {Cornish}, \citenamefont {Blas},\ and\ \citenamefont
  {Nardini}}]{Cornish:2017jml}%
  \BibitemOpen
  \bibfield  {author} {\bibinfo {author} {\bibfnamefont {N.}~\bibnamefont
  {Cornish}}, \bibinfo {author} {\bibfnamefont {D.}~\bibnamefont {Blas}},\ and\
  \bibinfo {author} {\bibfnamefont {G.}~\bibnamefont {Nardini}},\ }\href
  {https://doi.org/10.1103/PhysRevLett.119.161102} {\bibfield  {journal}
  {\bibinfo  {journal} {Phys. Rev. Lett.}\ }\textbf {\bibinfo {volume} {119}},\
  \bibinfo {pages} {161102} (\bibinfo {year} {2017})},\ \Eprint
  {https://arxiv.org/abs/1707.06101} {arXiv:1707.06101 [gr-qc]} \BibitemShut
  {NoStop}%
\bibitem [{\citenamefont {Khan}\ \emph {et~al.}(2019)\citenamefont {Khan},
  \citenamefont {Chatziioannou}, \citenamefont {Hannam},\ and\ \citenamefont
  {Ohme}}]{Khan_2019}%
  \BibitemOpen
  \bibfield  {author} {\bibinfo {author} {\bibfnamefont {S.}~\bibnamefont
  {Khan}}, \bibinfo {author} {\bibfnamefont {K.}~\bibnamefont {Chatziioannou}},
  \bibinfo {author} {\bibfnamefont {M.}~\bibnamefont {Hannam}},\ and\ \bibinfo
  {author} {\bibfnamefont {F.}~\bibnamefont {Ohme}},\ }\bibfield  {journal}
  {\bibinfo  {journal} {Physical Review D}\ }\textbf {\bibinfo {volume}
  {100}},\ \href {https://doi.org/10.1103/physrevd.100.024059}
  {10.1103/physrevd.100.024059} (\bibinfo {year} {2019})\BibitemShut {NoStop}%
\bibitem [{\citenamefont {Sana}\ \emph {et~al.}(2012)\citenamefont {Sana},
  \citenamefont {de~Mink}, \citenamefont {de~Koter}, \citenamefont {Langer},
  \citenamefont {Evans}, \citenamefont {Gieles}, \citenamefont {Gosset},
  \citenamefont {Izzard}, \citenamefont {Bouquin},\ and\ \citenamefont
  {Schneider}}]{Sana:2012px}%
  \BibitemOpen
  \bibfield  {author} {\bibinfo {author} {\bibfnamefont {H.}~\bibnamefont
  {Sana}}, \bibinfo {author} {\bibfnamefont {S.~E.}\ \bibnamefont {de~Mink}},
  \bibinfo {author} {\bibfnamefont {A.}~\bibnamefont {de~Koter}}, \bibinfo
  {author} {\bibfnamefont {N.}~\bibnamefont {Langer}}, \bibinfo {author}
  {\bibfnamefont {C.~J.}\ \bibnamefont {Evans}}, \bibinfo {author}
  {\bibfnamefont {M.}~\bibnamefont {Gieles}}, \bibinfo {author} {\bibfnamefont
  {E.}~\bibnamefont {Gosset}}, \bibinfo {author} {\bibfnamefont {R.~G.}\
  \bibnamefont {Izzard}}, \bibinfo {author} {\bibfnamefont {J.~B.~L.}\
  \bibnamefont {Bouquin}},\ and\ \bibinfo {author} {\bibfnamefont {F.~R.~N.}\
  \bibnamefont {Schneider}},\ }\href {https://doi.org/10.1126/science.1223344}
  {\bibfield  {journal} {\bibinfo  {journal} {Science}\ }\textbf {\bibinfo
  {volume} {337}},\ \bibinfo {pages} {444} (\bibinfo {year} {2012})},\ \Eprint
  {https://arxiv.org/abs/1207.6397} {arXiv:1207.6397 [astro-ph.SR]}
  \BibitemShut {NoStop}%
\bibitem [{\citenamefont {Abbott}\ \emph {et~al.}(2016)\citenamefont {Abbott}
  \emph {et~al.}}]{LIGOScientific:2016aoc}%
  \BibitemOpen
  \bibfield  {author} {\bibinfo {author} {\bibfnamefont {B.~P.}\ \bibnamefont
  {Abbott}} \emph {et~al.} (\bibinfo {collaboration} {LIGO Scientific,
  Virgo}),\ }\href {https://doi.org/10.1103/PhysRevLett.116.061102} {\bibfield
  {journal} {\bibinfo  {journal} {Phys. Rev. Lett.}\ }\textbf {\bibinfo
  {volume} {116}},\ \bibinfo {pages} {061102} (\bibinfo {year} {2016})},\
  \Eprint {https://arxiv.org/abs/1602.03837} {arXiv:1602.03837 [gr-qc]}
  \BibitemShut {NoStop}%
\bibitem [{\citenamefont {Abbott}\ \emph
  {et~al.}(2017{\natexlab{c}})\citenamefont {Abbott} \emph
  {et~al.}}]{LIGOScientific:2017ync}%
  \BibitemOpen
  \bibfield  {author} {\bibinfo {author} {\bibfnamefont {B.~P.}\ \bibnamefont
  {Abbott}} \emph {et~al.} (\bibinfo {collaboration} {LIGO Scientific, Virgo,
  Fermi GBM, INTEGRAL, IceCube, AstroSat Cadmium Zinc Telluride Imager Team,
  IPN, Insight-Hxmt, ANTARES, Swift, AGILE Team, 1M2H Team, Dark Energy Camera
  GW-EM, DES, DLT40, GRAWITA, Fermi-LAT, ATCA, ASKAP, Las Cumbres Observatory
  Group, OzGrav, DWF (Deeper Wider Faster Program), AST3, CAASTRO, VINROUGE,
  MASTER, J-GEM, GROWTH, JAGWAR, CaltechNRAO, TTU-NRAO, NuSTAR, Pan-STARRS,
  MAXI Team, TZAC Consortium, KU, Nordic Optical Telescope, ePESSTO, GROND,
  Texas Tech University, SALT Group, TOROS, BOOTES, MWA, CALET, IKI-GW
  Follow-up, H.E.S.S., LOFAR, LWA, HAWC, Pierre Auger, ALMA, Euro VLBI Team, Pi
  of Sky, Chandra Team at McGill University, DFN, ATLAS Telescopes, High Time
  Resolution Universe Survey, RIMAS, RATIR, SKA South Africa/MeerKAT}),\ }\href
  {https://doi.org/10.3847/2041-8213/aa91c9} {\bibfield  {journal} {\bibinfo
  {journal} {Astrophys. J. Lett.}\ }\textbf {\bibinfo {volume} {848}},\
  \bibinfo {pages} {L12} (\bibinfo {year} {2017}{\natexlab{c}})},\ \Eprint
  {https://arxiv.org/abs/1710.05833} {arXiv:1710.05833 [astro-ph.HE]}
  \BibitemShut {NoStop}%
\bibitem [{\citenamefont {Ciolfi}\ and\ \citenamefont
  {Siegel}(2014)}]{Ciolfi_2015}%
  \BibitemOpen
  \bibfield  {author} {\bibinfo {author} {\bibfnamefont {R.}~\bibnamefont
  {Ciolfi}}\ and\ \bibinfo {author} {\bibfnamefont {D.~M.}\ \bibnamefont
  {Siegel}},\ }\href {https://doi.org/10.1088/2041-8205/798/2/L36} {\bibfield
  {journal} {\bibinfo  {journal} {The Astrophysical Journal Letters}\ }\textbf
  {\bibinfo {volume} {798}},\ \bibinfo {pages} {L36} (\bibinfo {year}
  {2014})}\BibitemShut {NoStop}%
\bibitem [{\citenamefont {{Rezzolla}}\ and\ \citenamefont
  {{Kumar}}(2015)}]{2015ApJ...802...95R}%
  \BibitemOpen
  \bibfield  {author} {\bibinfo {author} {\bibfnamefont {L.}~\bibnamefont
  {{Rezzolla}}}\ and\ \bibinfo {author} {\bibfnamefont {P.}~\bibnamefont
  {{Kumar}}},\ }\href {https://doi.org/10.1088/0004-637X/802/2/95} {\bibfield
  {journal} {\bibinfo  {journal} {\apj}\ }\textbf {\bibinfo {volume} {802}},\
  \bibinfo {eid} {95} (\bibinfo {year} {2015})},\ \Eprint
  {https://arxiv.org/abs/1410.8560} {arXiv:1410.8560 [astro-ph.HE]}
  \BibitemShut {NoStop}%
\bibitem [{\citenamefont {Abadie}\ \emph {et~al.}(2012)\citenamefont {Abadie}
  \emph {et~al.}}]{LIGOScientific:2012fcp}%
  \BibitemOpen
  \bibfield  {author} {\bibinfo {author} {\bibfnamefont {J.}~\bibnamefont
  {Abadie}} \emph {et~al.} (\bibinfo {collaboration} {LIGO Scientific}),\
  }\href {https://doi.org/10.1088/0004-637X/760/1/12} {\bibfield  {journal}
  {\bibinfo  {journal} {Astrophys. J.}\ }\textbf {\bibinfo {volume} {760}},\
  \bibinfo {pages} {12} (\bibinfo {year} {2012})},\ \Eprint
  {https://arxiv.org/abs/1205.2216} {arXiv:1205.2216 [astro-ph.HE]}
  \BibitemShut {NoStop}%
\bibitem [{\citenamefont {Finn}\ \emph {et~al.}(1999)\citenamefont {Finn},
  \citenamefont {Mohanty},\ and\ \citenamefont {Romano}}]{Finn:1999vh}%
  \BibitemOpen
  \bibfield  {author} {\bibinfo {author} {\bibfnamefont {L.~S.}\ \bibnamefont
  {Finn}}, \bibinfo {author} {\bibfnamefont {S.~D.}\ \bibnamefont {Mohanty}},\
  and\ \bibinfo {author} {\bibfnamefont {J.~D.}\ \bibnamefont {Romano}},\
  }\href {https://doi.org/10.1103/PhysRevD.60.121101} {\bibfield  {journal}
  {\bibinfo  {journal} {Phys. Rev. D}\ }\textbf {\bibinfo {volume} {60}},\
  \bibinfo {pages} {121101} (\bibinfo {year} {1999})},\ \Eprint
  {https://arxiv.org/abs/gr-qc/9903101} {arXiv:gr-qc/9903101} \BibitemShut
  {NoStop}%
\bibitem [{\citenamefont {Coulter}\ \emph {et~al.}(2017)\citenamefont
  {Coulter}, \citenamefont {Foley}, \citenamefont {Kilpatrick}, \citenamefont
  {Drout}, \citenamefont {Piro}, \citenamefont {Shappee}, \citenamefont
  {Siebert}, \citenamefont {Simon}, \citenamefont {Ulloa}, \citenamefont
  {Kasen}, \citenamefont {Madore}, \citenamefont {Murguia-Berthier},
  \citenamefont {Pan}, \citenamefont {Prochaska}, \citenamefont {Ramirez-Ruiz},
  \citenamefont {Rest},\ and\ \citenamefont
  {Rojas-Bravo}}]{doi:10.1126/science.aap9811}%
  \BibitemOpen
  \bibfield  {author} {\bibinfo {author} {\bibfnamefont {D.~A.}\ \bibnamefont
  {Coulter}}, \bibinfo {author} {\bibfnamefont {R.~J.}\ \bibnamefont {Foley}},
  \bibinfo {author} {\bibfnamefont {C.~D.}\ \bibnamefont {Kilpatrick}},
  \bibinfo {author} {\bibfnamefont {M.~R.}\ \bibnamefont {Drout}}, \bibinfo
  {author} {\bibfnamefont {A.~L.}\ \bibnamefont {Piro}}, \bibinfo {author}
  {\bibfnamefont {B.~J.}\ \bibnamefont {Shappee}}, \bibinfo {author}
  {\bibfnamefont {M.~R.}\ \bibnamefont {Siebert}}, \bibinfo {author}
  {\bibfnamefont {J.~D.}\ \bibnamefont {Simon}}, \bibinfo {author}
  {\bibfnamefont {N.}~\bibnamefont {Ulloa}}, \bibinfo {author} {\bibfnamefont
  {D.}~\bibnamefont {Kasen}}, \bibinfo {author} {\bibfnamefont {B.~F.}\
  \bibnamefont {Madore}}, \bibinfo {author} {\bibfnamefont {A.}~\bibnamefont
  {Murguia-Berthier}}, \bibinfo {author} {\bibfnamefont {Y.-C.}\ \bibnamefont
  {Pan}}, \bibinfo {author} {\bibfnamefont {J.~X.}\ \bibnamefont {Prochaska}},
  \bibinfo {author} {\bibfnamefont {E.}~\bibnamefont {Ramirez-Ruiz}}, \bibinfo
  {author} {\bibfnamefont {A.}~\bibnamefont {Rest}},\ and\ \bibinfo {author}
  {\bibfnamefont {C.}~\bibnamefont {Rojas-Bravo}},\ }\href
  {https://doi.org/10.1126/science.aap9811} {\bibfield  {journal} {\bibinfo
  {journal} {Science}\ }\textbf {\bibinfo {volume} {358}},\ \bibinfo {pages}
  {1556} (\bibinfo {year} {2017})},\ \Eprint
  {https://arxiv.org/abs/https://www.science.org/doi/pdf/10.1126/science.aap9811}
  {https://www.science.org/doi/pdf/10.1126/science.aap9811} \BibitemShut
  {NoStop}%
\bibitem [{\citenamefont {Goldreich}\ and\ \citenamefont
  {Julian}(1969)}]{Goldreich:1969sb}%
  \BibitemOpen
  \bibfield  {author} {\bibinfo {author} {\bibfnamefont {P.}~\bibnamefont
  {Goldreich}}\ and\ \bibinfo {author} {\bibfnamefont {W.~H.}\ \bibnamefont
  {Julian}},\ }\href {https://doi.org/10.1086/150119} {\bibfield  {journal}
  {\bibinfo  {journal} {Astrophys. J.}\ }\textbf {\bibinfo {volume} {157}},\
  \bibinfo {pages} {869} (\bibinfo {year} {1969})}\BibitemShut {NoStop}%
\bibitem [{\citenamefont {Contopoulos}\ \emph {et~al.}(1999)\citenamefont
  {Contopoulos}, \citenamefont {Kazanas},\ and\ \citenamefont
  {Fendt}}]{Contopoulos_1999}%
  \BibitemOpen
  \bibfield  {author} {\bibinfo {author} {\bibfnamefont {I.}~\bibnamefont
  {Contopoulos}}, \bibinfo {author} {\bibfnamefont {D.}~\bibnamefont
  {Kazanas}},\ and\ \bibinfo {author} {\bibfnamefont {C.}~\bibnamefont
  {Fendt}},\ }\href {https://doi.org/10.1086/306652} {\bibfield  {journal}
  {\bibinfo  {journal} {The Astrophysical Journal}\ }\textbf {\bibinfo {volume}
  {511}},\ \bibinfo {pages} {351} (\bibinfo {year} {1999})}\BibitemShut
  {NoStop}%
\bibitem [{\citenamefont {Spitkovsky}(2006)}]{Spitkovsky_2006}%
  \BibitemOpen
  \bibfield  {author} {\bibinfo {author} {\bibfnamefont {A.}~\bibnamefont
  {Spitkovsky}},\ }\href {https://doi.org/10.1086/507518} {\bibfield  {journal}
  {\bibinfo  {journal} {The Astrophysical Journal}\ }\textbf {\bibinfo {volume}
  {648}},\ \bibinfo {pages} {L51} (\bibinfo {year} {2006})}\BibitemShut
  {NoStop}%
\bibitem [{\citenamefont {Castellanos}\ and\ \citenamefont
  {Chacon-Acosta}(2013)}]{Castellanos:2013ru}%
  \BibitemOpen
  \bibfield  {author} {\bibinfo {author} {\bibfnamefont {E.}~\bibnamefont
  {Castellanos}}\ and\ \bibinfo {author} {\bibfnamefont {G.}~\bibnamefont
  {Chacon-Acosta}},\ }\href {https://doi.org/10.1016/j.physletb.2013.04.009}
  {\bibfield  {journal} {\bibinfo  {journal} {Phys. Lett. B}\ }\textbf
  {\bibinfo {volume} {722}},\ \bibinfo {pages} {119} (\bibinfo {year}
  {2013})},\ \Eprint {https://arxiv.org/abs/1301.5362} {arXiv:1301.5362
  [gr-qc]} \BibitemShut {NoStop}%
\bibitem [{\citenamefont {Chac\'on-Acosta}\ and\ \citenamefont
  {Hernandez-Hernandez}(2015)}]{Chacon-Acosta:2014zva}%
  \BibitemOpen
  \bibfield  {author} {\bibinfo {author} {\bibfnamefont {G.}~\bibnamefont
  {Chac\'on-Acosta}}\ and\ \bibinfo {author} {\bibfnamefont {H.~H.}\
  \bibnamefont {Hernandez-Hernandez}},\ }\href
  {https://doi.org/10.1142/S0218271815500339} {\bibfield  {journal} {\bibinfo
  {journal} {Int. J. Mod. Phys. D}\ }\textbf {\bibinfo {volume} {24}},\
  \bibinfo {pages} {1550033} (\bibinfo {year} {2015})},\ \Eprint
  {https://arxiv.org/abs/1408.1306} {arXiv:1408.1306 [astro-ph.SR]}
  \BibitemShut {NoStop}%
\bibitem [{\citenamefont {Demir}\ and\ \citenamefont
  {Sargin}(2014)}]{Demir:2014rfa}%
  \BibitemOpen
  \bibfield  {author} {\bibinfo {author} {\bibfnamefont {D.~A.}\ \bibnamefont
  {Demir}}\ and\ \bibinfo {author} {\bibfnamefont {O.}~\bibnamefont {Sargin}},\
  }\href {https://doi.org/10.1016/j.physleta.2014.09.044} {\bibfield  {journal}
  {\bibinfo  {journal} {Phys. Lett. A}\ }\textbf {\bibinfo {volume} {378}},\
  \bibinfo {pages} {3237} (\bibinfo {year} {2014})},\ \Eprint
  {https://arxiv.org/abs/1409.7224} {arXiv:1409.7224 [quant-ph]} \BibitemShut
  {NoStop}%
\bibitem [{\citenamefont {Demarie}\ and\ \citenamefont
  {Terno}(2013)}]{Demarie:2012tz}%
  \BibitemOpen
  \bibfield  {author} {\bibinfo {author} {\bibfnamefont {T.~F.}\ \bibnamefont
  {Demarie}}\ and\ \bibinfo {author} {\bibfnamefont {D.~R.}\ \bibnamefont
  {Terno}},\ }\href {https://doi.org/10.1088/0264-9381/30/13/135006} {\bibfield
   {journal} {\bibinfo  {journal} {Class. Quant. Grav.}\ }\textbf {\bibinfo
  {volume} {30}},\ \bibinfo {pages} {135006} (\bibinfo {year} {2013})},\
  \Eprint {https://arxiv.org/abs/1209.3087} {arXiv:1209.3087 [gr-qc]}
  \BibitemShut {NoStop}%
\bibitem [{\citenamefont {Nozari}\ \emph {et~al.}(2015)\citenamefont {Nozari},
  \citenamefont {Khodadi},\ and\ \citenamefont {Gorji}}]{Nozari:2015qoi}%
  \BibitemOpen
  \bibfield  {author} {\bibinfo {author} {\bibfnamefont {K.}~\bibnamefont
  {Nozari}}, \bibinfo {author} {\bibfnamefont {M.}~\bibnamefont {Khodadi}},\
  and\ \bibinfo {author} {\bibfnamefont {M.~A.}\ \bibnamefont {Gorji}},\ }\href
  {https://doi.org/10.1209/0295-5075/112/60003} {\bibfield  {journal} {\bibinfo
   {journal} {EPL}\ }\textbf {\bibinfo {volume} {112}},\ \bibinfo {pages}
  {60003} (\bibinfo {year} {2015})},\ \Eprint
  {https://arxiv.org/abs/1512.07779} {arXiv:1512.07779 [gr-qc]} \BibitemShut
  {NoStop}%
\bibitem [{\citenamefont {Khodadi}\ \emph {et~al.}(2018)\citenamefont
  {Khodadi}, \citenamefont {Nozari}, \citenamefont {Dey}, \citenamefont
  {Bhat},\ and\ \citenamefont {Faizal}}]{Khodadi:2017kah}%
  \BibitemOpen
  \bibfield  {author} {\bibinfo {author} {\bibfnamefont {M.}~\bibnamefont
  {Khodadi}}, \bibinfo {author} {\bibfnamefont {K.}~\bibnamefont {Nozari}},
  \bibinfo {author} {\bibfnamefont {S.}~\bibnamefont {Dey}}, \bibinfo {author}
  {\bibfnamefont {A.}~\bibnamefont {Bhat}},\ and\ \bibinfo {author}
  {\bibfnamefont {M.}~\bibnamefont {Faizal}},\ }\href
  {https://doi.org/10.1038/s41598-018-19181-9} {\bibfield  {journal} {\bibinfo
  {journal} {Sci. Rep.}\ }\textbf {\bibinfo {volume} {8}},\ \bibinfo {pages}
  {1659} (\bibinfo {year} {2018})},\ \Eprint {https://arxiv.org/abs/1801.00273}
  {arXiv:1801.00273 [gr-qc]} \BibitemShut {NoStop}%
\bibitem [{\citenamefont {Abbott}\ \emph
  {et~al.}(2021{\natexlab{a}})\citenamefont {Abbott} \emph
  {et~al.}}]{LIGOScientific:2020ibl}%
  \BibitemOpen
  \bibfield  {author} {\bibinfo {author} {\bibfnamefont {R.}~\bibnamefont
  {Abbott}} \emph {et~al.} (\bibinfo {collaboration} {LIGO Scientific,
  Virgo}),\ }\href {https://doi.org/10.1103/PhysRevX.11.021053} {\bibfield
  {journal} {\bibinfo  {journal} {Phys. Rev. X}\ }\textbf {\bibinfo {volume}
  {11}},\ \bibinfo {pages} {021053} (\bibinfo {year} {2021}{\natexlab{a}})},\
  \Eprint {https://arxiv.org/abs/2010.14527} {arXiv:2010.14527 [gr-qc]}
  \BibitemShut {NoStop}%
\bibitem [{\citenamefont {Abbott}\ \emph
  {et~al.}(2021{\natexlab{b}})\citenamefont {Abbott} \emph
  {et~al.}}]{LIGOScientific:2021usb}%
  \BibitemOpen
  \bibfield  {author} {\bibinfo {author} {\bibfnamefont {R.}~\bibnamefont
  {Abbott}} \emph {et~al.} (\bibinfo {collaboration} {LIGO Scientific,
  VIRGO}),\ }\href@noop {} {\  (\bibinfo {year} {2021}{\natexlab{b}})},\
  \Eprint {https://arxiv.org/abs/2108.01045} {arXiv:2108.01045 [gr-qc]}
  \BibitemShut {NoStop}%
\bibitem [{\citenamefont {Abbott}\ \emph
  {et~al.}(2021{\natexlab{c}})\citenamefont {Abbott} \emph
  {et~al.}}]{LIGOScientific:2021djp}%
  \BibitemOpen
  \bibfield  {author} {\bibinfo {author} {\bibfnamefont {R.}~\bibnamefont
  {Abbott}} \emph {et~al.} (\bibinfo {collaboration} {LIGO Scientific, VIRGO,
  KAGRA}),\ }\href@noop {} {\  (\bibinfo {year} {2021}{\natexlab{c}})},\
  \Eprint {https://arxiv.org/abs/2111.03606} {arXiv:2111.03606 [gr-qc]}
  \BibitemShut {NoStop}%
\bibitem [{\citenamefont {Iacovelli}\ \emph {et~al.}(2022)\citenamefont
  {Iacovelli}, \citenamefont {Mancarella}, \citenamefont {Foffa},\ and\
  \citenamefont {Maggiore}}]{Iacovelli_2022}%
  \BibitemOpen
  \bibfield  {author} {\bibinfo {author} {\bibfnamefont {F.}~\bibnamefont
  {Iacovelli}}, \bibinfo {author} {\bibfnamefont {M.}~\bibnamefont
  {Mancarella}}, \bibinfo {author} {\bibfnamefont {S.}~\bibnamefont {Foffa}},\
  and\ \bibinfo {author} {\bibfnamefont {M.}~\bibnamefont {Maggiore}},\ }\href
  {https://doi.org/10.3847/1538-4357/ac9cd4} {\bibfield  {journal} {\bibinfo
  {journal} {The Astrophysical Journal}\ }\textbf {\bibinfo {volume} {941}},\
  \bibinfo {pages} {208} (\bibinfo {year} {2022})}\BibitemShut {NoStop}%
\end{thebibliography}%

\end{document}